\documentclass[useAMS, usenatbib]{mn2e}
\usepackage{amssymb,amsmath}
\usepackage{graphicx}

\title{Time Dependent Radiative Transfer for Multi-Level Atoms using 
       Accelerated Lambda Iteration}
\author[Matthew van Adelsberg and Rosalba Perna]{Matthew van 
Adelsberg$^{1}$\thanks{Email: 
miva3@mail.gatech.edu} and Rosalba Perna$^{2}$\\
$^{1}$Centre for Relativistic Astrophysics, Georgia Institute of Technology, 
Atlanta, GA 30332, USA\\
$^{2}$JILA and Department of Astrophysical and Planetary Science, 
University of Colorado at Boulder, Boulder, CO 80304, USA}

\begin{document}

\date{Accepted --.  Received --; in original form --}
\pagerange{\pageref{firstpage}--\pageref{lastpage}} \pubyear{2012}
\maketitle

\label{firstpage}

\begin{abstract}
We present a general formalism for computing self-consistent, numerical 
solutions to the time-dependent radiative transfer equation in low 
velocity, multi-level ions undergoing radiative interactions.
Recent studies of time-dependent radiative transfer have focused on 
radiation hydrodynamic and magnetohydrodynamic effects without 
lines, or have solved time-independent equations for the radiation field 
simultaneously with time-dependent equations for the state of the medium.
In this paper, we provide a fully time-dependent numerical
solution to the radiative transfer and atomic rate equations for a 
medium irradiated by an external source of photons.  
We use Accelerated 
Lambda Iteration to achieve convergence of the radiation field 
and atomic states.  We perform calculations 
for a three-level atomic model that illustrates important 
time-dependent effects.  We demonstrate that our 
method provides an efficient, accurate solution to the time-dependent radiative 
transfer problem.  Finally, we characterize astrophysical 
scenarios in which we expect our solutions to be important. 
\end{abstract}

\begin{keywords}
line: formation -- radiative transfer -- scattering -- gamma-rays: bursts -- galaxies: active
\end{keywords}

\section{Introduction}
\label{sect:Introduction}

The treatment of radiation transfer in complex, dynamic physical
systems is crucial for theoretical modelling in a wide variety of
astrophysical contexts.  In this paper, we explore time-dependent
effects in line transfer for low velocity media 
illuminated by external sources.  
Examples of such systems are found in absorbing material 
irradiated by light from gamma-ray bursts (GRBs) and active
galactic nuclei (AGNs). 
In absorbing media, external photons enter at
the boundary of the system, and are subsequently redistributed in
angle and frequency by atomic interactions in the interior.  This
leads to a complicated, time-dependent coupling of the radiation field 
and the state of the medium.

Previous studies have investigated time-dependent
radiative transfer in a number of regimes, under different
approximations.  Several papers have explored steady-state, 
non-LTE solutions to the radiative transfer equation (RTE) 
for time-dependent media.  These works were done in the contexts of
magnetohydrodynamics \citep[e.g.,][]{Hayeketal10a, Davisetal12a},
radiation hydrodynamics \citep[e.g.,][]{Krumholzetal12a}, 
smoothed particle hydrodynamics \citep[e.g.,][]{Rundleetal10a}, transport of
cosmological ionization fronts \citep[e.g.,][]{WhalenNorman06a}, molecular bands
in planetary atmospheres \citep[e.g.,][]{Kutepovetal91a}, line transfer in
stellar atmospheres \citep[e.g.,][]{Sellmaieretal93a}, multi-level transfer 
in moving media \citep[e.g.,][]{Hauschildt93a, Harper94a}, atmospheres of hot stars
\citep[e.g.,][]{HubenyLanz95a}, clumpy molecular clouds \citep[e.g.,][]{Juvela97a},
and line emission from interstellar clouds \citep[e.g.,][]{JuvelaPadoan05a}.

A few authors have investigated solutions to the time-dependent
radiative transfer equation (TDRTE). \citet[][]{PernaLazzati02a} and
\citet[][]{Pernaetal03a} treated time-dependent radiative transfer in 
static, dusty media in the optically thin limit, without 
radiation source terms.
\citealt[][]{GnedinAbel01a} and \citealt[][]{AbelWandelt02a} 
also solved the time-dependent radiative transfer equation 
for cosmological reionization, neglecting photon scattering. 
\citet[][]{MihalasKlein82a} developed a `mixed-frame' formalism for
solving the time-dependent equations in the context of fluid flow, 
and investigated numerical
methods for their solution.  \citet[][]{Kunasz83a} calculated the
numerical solution to the time-dependent transfer equation for resonance 
scattering in a static medium consisting of two-level atoms.  

Recently,
\citet[][]{HubenyBurrows07a} produced a self-consistent solution to
the TDRTE for a time-dependent, non-LTE medium in 
radiation hydrodynamics.  Their calculation focused on the propagation
of neutrinos in supernova explosions, and did not explicitly treat
line transfer of photons \citep[see also][for further discussion of 
time-dependent neutrino radiation transport]{Abdikamalovetal12a}.

In this paper, we neglect fluid flow, and focus instead on 
time-dependent effects in the line transfer of photons.
In externally driven absorbing material, the time derivative of the 
RTE cannot be neglected due to 
the variability of the incoming radiation at the boundary of the system.  
The resulting solution depends on several crucial timescales: the 
variability timescale 
of the external source, $T_{\rm VR}$; the time interval between 
characteristic radiative interactions, $T_{\rm ref}$; and the light travel 
time across the width of the medium, $T_{\rm LC}$.  
For optically thick systems, $T_{\rm ref} \ll T_{\rm LC}$.
If $T_{\rm VR}\ll T_{\rm ref}$, we argue that the source term for atomic 
interactions is unimportant for solving the TDRTE, and a simple, 
approximate solution is appropriate.  Conversely, if 
$T_{\rm VR}\gg T_{\rm LC}$, 
then the RTE can be solved in steady-state, while the medium is treated as 
time-dependent; this situation has been discussed exhaustively 
in the literature.  
Thus, we are primarily interested in the `intermediate regime',
$T_{\rm VR} \gtrsim T_{\rm ref}$, $T_{\rm VR}\lesssim T_{\rm LC}$, and will perform 
a detailed analysis of this scenario in the work that follows.

The solution to the non-LTE radiative transfer problem has 
an enormous associated literature, a proper review of which is beyond the 
scope of this paper.   In the work that follows, we extend the Accelerated 
Lambda Iteration (ALI) method to the case of time-dependent 
radiative transfer and atomic rate equations.
The ALI approach employs a computationally inexpensive 
approximation to the RTE formal solution in the rate equations for the 
atomic level populations.  This technique has its origins in the `core 
saturation' method of \citet[][]{Rybicki72a, Rybicki84a}, and Cannon's 
operator
splitting procedure \citep[][]{Cannon73a}.  \citet[][]{Olsonetal86a} showed 
that the diagonal part of the operator characterizing the formal solution 
to the RTE provides an efficient approximate operator for implementing 
ALI.  Subsequently, \citet[][]{RybickiHummer91a, RybickiHummer92a} developed 
a powerful formalism for applying ALI to non-LTE problems with multi-level 
atomic models.  For a detailed review of ALI, see \citet[][]{Hubeny03a}.

In this paper, we extend the TDRTE solution of \citet[][]{Kunasz83a} to
time-dependent media with multi-level atoms, making extensive use of the 
\citet[][]{RybickiHummer92a} formalism.  To date, we have found no previous 
work in the literature that treats both the RTE and atomic rate equations 
as time variable for multi-level line transfer.
Our method applies to general atomic models, including
bound-bound and bound-free radiative transitions, 
collisional interactions, electron scattering, and background processes such as 
free-free absorption and emission.
We present calculations for a specific three-level atomic model that 
illustrates important time-dependent radiative transfer effects. 
We find that these effects are important 
for systems in the intermediate regime and show that the use 
of approximate solutions that neglect them can lead to significant 
errors when interpreting spectroscopic observations.  
We consider the implications of our results for astrophysical systems, and 
provide a set of criteria for determining when the use of our formalism is 
necessary.
 
Our paper has the following outline: in \S\ref{sect:PhysModel}, we describe our 
models for the atomic physics and external radiation sources used in our 
calculations; in \S\ref{sect:Equations}, we review the basic equations for 
the TDRTE and atomic rate equations; in \S\ref{sect:Numerical}, 
we develop a general formalism for implementing the ALI solution; 
in \S\ref{sect:Results}, we show the results of our calculations; and in 
\S\ref{sect:Discussion}, we discuss implications of our results for 
astrophysical systems.  
In Appendix~\ref{sect:RKMethods}, we extend our 
technique to include implicit Runge-Kutta time integration of our equations.

\section{Models for Atomic Physics and External Radiation}
\label{sect:PhysModel}

We consider a plane parallel, finite slab of material that is illuminated 
by an external source of unpolarized radiation.  The properties of the medium 
vary along the $z$ axis.  The cosine of the polar angle, $\theta$, measured 
from the $z$ axis, is denoted by $\mu$.  The system is assumed to be symmetric 
in the azimuthal direction.

The external radiation propagates toward the slab with  
$\theta > \pi / 2$, reaching the medium at an arbitrary coordinate, 
$z_{\rm o}$.  The width of the slab along the $z$ direction is denoted 
by $W$; thus, radiation exits the slab at $z_{\rm min} = z_{\rm o} - W$. 
After interacting with the medium, the radiation field in the slab 
consists of rays with $-\pi\le \theta\le \pi$.  
We assume that an observer is positioned in the 
$\theta = \pi$ direction.

To illustrate key properties of the time-dependent solutions, we 
perform calculations for a specific three-level atomic model.  
The atomic levels are labelled by $j = 0, 1, 2$, and are connected by 
bound-bound radiative transitions between $j = 0\leftrightarrow 1$, 
$j = 0\leftrightarrow 2$, and $j = 1\leftrightarrow 2$.  We neglect 
other types of 
radiative and collisional interactions in our calculations, though 
they are 
straightforwardly included in our formalism, which is completely general.

The number density of the atoms is denoted by $n_{\rm o}$.  We 
define a reference transition for photoexcitation between the atomic levels 
$j = 0\rightarrow 2$, at line centre, with all atoms in the ground state.  
If the Einstein coefficient for this 
transition is $B_{02}$, and the line profile is due to Doppler broadening
in complete redistribution, we obtain a reference extinction coefficient,
\begin{equation}
\chi_{\rm ref} = \frac{h B_{02} n_{\rm o}}{4\pi^{3/2} b /c},
\end{equation}
where $b = \sqrt{2 k_B T_a / m_a}$ is the Doppler width for an 
atom with temperature $T_a$ and mass $m_a$ (ignoring 
possible effects of microturbulence).
We define a reference optical depth,
\begin{equation}
d\tau \equiv - \chi_{\rm ref} dz,
\end{equation}
such that $\tau(z_{\rm o}) = 0$, and the maximum optical depth is 
$\tau(z_{\rm min})\equiv \tau_{\rm max} = \chi_{\rm ref} W$.
We note that our definitions of $z$ and $\tau$ are opposite to 
those used in the stellar atmospheres literature, in which the 
optical depth increases with distance from the observer.

If we define a reference time,
\begin{equation}
T_{\rm ref} \equiv \left(c \chi_{\rm ref}\right)^{-1},
\end{equation}
then the light crossing time for radiation passing through the slab is 
\begin{equation}
T_{\rm LC} = W / c = \tau_{\rm max} T_{\rm ref}.
\end{equation}
A natural unit of radiation intensity can be defined by equating the 
inverse of the photoexcitation rate for white, isotropic radiation 
exciting the reference transition, to the 
reference time:
\begin{equation}
I_{\rm o} \equiv c \chi_{\rm ref} / B_{02}.
\end{equation}
When the photoexcitation rate for the reference transition is equal to 
$B_{02} I_{\rm o}$, the number density of line centre photons is 
equal to $n_{\rm o}$.

The Einstein coefficients for 
the transitions in our atomic model, using the units described above, are 
summarized 
in Table~\ref{tab:AtomicData}.
\begin{table}
\caption{Einstein coefficients for the transitions in our atomic model.  $A$, 
$B_{\rm ex}$, and $B_{\rm em}$ 
are the coefficients for spontaneous emission, photoexcitation, and stimulated 
emission, respectively.}
\centering
\begin{tabular}{c c c c}
Transition & $A / T_{\rm ref}^{-1}$ & $B_{\rm ex} / B_{02}$
	   & $B_{\rm em} / B_{02}$\\
\hline
$0\rightarrow 2$ & $1.4\times 10^{10}$  & 1 & 1.25\\ \hline
$1\rightarrow 2$ & $4.9\times 10^9$ & 0.44 & 0.44\\ \hline
$0\rightarrow 1$ & 0.18 & $1.7 \times 10^{-5}$ & $2.1\times 10^{-5}$\\ \hline
\end{tabular}
\label{tab:AtomicData}
\end{table}
The values of the coefficients have been chosen so that the atomic model 
contains a resonance line ($j = 0\leftrightarrow 2$), a strong transition 
between the excited states ($j = 1\leftrightarrow 2$), and a weak 
transition between the ground and first excited states 
($j = 0\leftrightarrow 1$). 
This arrangement insures that the second excited state ($j = 2$) remains 
relatively 
unpopulated, while the transition $j = 0\rightarrow 2\rightarrow 1$ provides 
an efficient mechanism for populating the first excited 
state.\footnote{The atomic 
data in Table~\ref{tab:AtomicData} was chosen to mimic 
transitions from the ground state ($^6D_{9/2}$) and first excited 
state ($^6D_{7/2}$) to a common upper state ($^6P_{7/2}$) of the Fe~II 
ion in fine structure splitting.  
Nevertheless, our 
three-level model provides a simple, general mechanism for populating an 
excited state by a resonant transition; this mechanism is widely 
applicable to many ions.}

In \S\ref{sect:Numerical}, to make comparisons to the previous work 
of \citet[][]{Kunasz83a},
we also consider a restricted version of the atomic model that includes 
only the single transition $j= 0\rightarrow 2$.

For the calculations in this paper, we consider several basic models for the 
external radiation arriving at $\tau = 0$.
We assume that the angular variation of the incident radiation is 
either isotropic or highly collimated at the $\tau = 0$ boundary of the slab.  
In the latter case, photons arrive from the external source 
in a narrow range of angles around the line of sight.
In both cases, the spectrum of the external radiation is assumed to be 
white across all three atomic transitions.

We employ a heuristic model for the time variation of the arriving photons.
This model combines a transient increase in intensity with a constant  
background intensity:
\begin{equation}
\label{eq:lightcurve}
I_{\nu}^{\rm ext}(t, \mu) = \left\{I_{\rm B} + 
   A \exp\left[-\frac{1}{2}\log^2\left(
   t / t_0\right) / \sigma^2\right]\right\} f(\mu),
\end{equation}
where $I_{\nu}^{\rm ext}$ is the specific intensity of the incident  
radiation at $\tau = 0$, $I_{\rm B}$ is a constant background intensity, 
and $A$ is the amplitude of the transient part of the incoming 
radiation.  The variable part of 
equation~\eqref{eq:lightcurve} is a Gaussian pulse in $\log t$, where 
the parameters $t_0$ and $\sigma$ determine the centring and width of the 
pulse, respectively.  The angular variation $f(\mu)$ is set to unity for 
isotropic sources, and takes the value $f(\mu) = \delta(\mu + 1)$ for 
collimated sources.  Thus, if $A\gg I_{\rm B}$, the external source emits 
photons in a transient pulse, whereas if $A\gtrsim I_{\rm B}$, the external 
radiation represents a persistent source with superimposed variability.

To make comparisons to the previous work of \citet[][]{Kunasz83a},
we also employ a simple `ramp' model in which the external source 
increases linearly  from 
zero at $t / T_{\rm ref} = 0$ to $I_{\rm B}$ over 
the time interval $10^{-2} T_{\rm ref}$, remaining constant thereafter.  

\section{Fundamental Equations}
\label{sect:Equations}

\subsection{Radiative Transfer Equation}
\label{subsect:RTE}

We restrict the range of angular values to $\mu\in [0,1]$, and write 
the specific intensity for the unpolarized radiation field as two 
distinct functions, $I_{\nu}^+(t, z, \mu)$ for
$\theta \le \pi / 2$ and $I_{\nu}^-(t, z, \mu)$ for $\theta > \pi/2$. 
With this notation, the TDRTE can be written:
\begin{equation}
\label{eq:TDRTE}
\frac{1}{c}\frac{\partial I_{\nu}^{\pm}}{\partial t} \pm 
    \mu \frac{\partial I_{\nu}^{\pm}}{\partial z} = -\chi_{\nu} I_{\nu}^{\pm}
    + \eta_{\nu}.
\end{equation}
The extinction and emission coefficients can be cast in a general form 
for bound-bound and bound-free radiative interactions 
\citep[c.f.,][]{RybickiHummer92a}.  We denote the 
frequency-angle rate coefficient for spontaneous emission or radiative 
recombination as $U_{lj}$, depending on whether 
$l\rightarrow j$ represents a bound-bound or bound-free transition.
Note that $U_{lj} = 0$ if $l \le j$. 
Similarly, $V_{lj}$ denotes stimulated emission or recombination for $l > j$, 
and absorption or photoionization for $l < j$.  For the familiar 
case of bound-bound transitions with complete redistribution in the line 
profile, the rate coefficients become:
\begin{align}
\label{eq:CRU}
U_{lj} & = 
    \left\{\begin{aligned}
        \frac{h\nu_{lj}}{4\pi} A_{lj} \varphi_{lj}, &\quad l > j\\
        0, &\quad l < j
    \end{aligned}\right.\\
\label{eq:CRV}
V_{lj} & = \frac{h\nu_{lj}}{4\pi} B_{lj} \varphi_{lj},
\end{align}
where $\nu_{lj}$ is the line centre frequency of the transition, and 
$\varphi_{lj}$ is the absorption line profile.  Note that under the 
assumption of complete redistribution, $\varphi_{lj} = \varphi_{jl}$.  
Similar equations can be 
written for bound-free interactions, or for different approximations for the 
redistribution of the line profile 
\citep[see, e.g.,][]{RybickiHummer92a, Uitenbroek01a}.  In this notation, 
the extinction and emission coefficients become:
\begin{align}
\label{eq:chi}
\chi_{\nu} & = \sum_{l, j;\, l < j} \chi_{lj} + \chi_{\rm c},\\
\label{eq:eta}
\eta_{\nu} & = \sum_{l, j} U_{lj} n_l + \eta_c,
\end{align}
where
\begin{equation}
\chi_{lj} \equiv V_{lj} n_l - V_{jl} n_j,
\end{equation}
and $\chi_c$, $\eta_c$ represent background processes or 
electron scattering; these contributions are treated separately in the 
numerical method from the bound-bound and bound-free interactions 
(see \S\ref{sect:Numerical}).  The $\Sigma$ symbols in 
equations~\eqref{eq:chi} and \eqref{eq:eta} denote double sums over the 
$j$ and $l$ indices for all transitions satisfying the indicated condition.
The number density of atoms in level $l$ is denoted by $n_l$.

Equation~\eqref{eq:TDRTE} must be supplemented by initial and boundary 
conditions for the specific intensity in each angular range.  The boundary 
conditions take the form:
\begin{gather}
\label{eq:RTEBC-}
I_{\nu}^-\left(t, z_{\rm o}, \mu\right) = I_{\nu}^{\rm ext}(t, \mu),\\
\label{eq:RTEBC+}
I_{\nu}^+\left(t, z_{\rm min}, \mu\right) = 0.
\end{gather} 
The initial conditions are determined by solving the time independent 
radiative transfer equation (TIRTE), which is obtained by setting 
$\partial I_{\nu}^{\pm}/\partial t \rightarrow 0$ 
in equation~\eqref{eq:TDRTE}.
Equations~\eqref{eq:RTEBC-} and \eqref{eq:RTEBC+}, evaluated at 
$t / T_{\rm ref}= 0$, are then used as boundary conditions for the TIRTE.  
The resulting 
solution is used for the initial conditions, $I_{\nu}^{\pm}(t = 0, z, \mu)$.
The TIRTE can be solved with an approach that is similar to 
the time-dependent method (see \S\ref{sect:Numerical}).

In developing numerical methods for solving the RTE, it is useful to define 
the Feautrier variables 
\citep[see, e.g.,][and the references therein]{Mihalas78a}:
\begin{align}
\label{eq:Feautu}
u_{\nu} & = \left(I_{\nu}^+ + I_{\nu}^-\right)/2,\\
\label{eq:Feautv}
v_{\nu} & = \left(I_{\nu}^+ - I_{\nu}^-\right)/2.
\end{align}
We define the optical depth along a ray path as:
\begin{equation}
d\tau_{\nu} = -\chi_{\nu} dz / \mu,
\end{equation}
where $\tau_{\nu}(z_{\rm o}) = 0$.
A change in variables to $u_{\nu}$, $v_{\nu}$, and $\tau_{\nu}$, 
results in the following form for the transfer equations:
\begin{align}
\label{eq:FeautRTEu}
\frac{\partial u_{\nu}}{\partial t} & = c\chi_{\nu}\left(
    \frac{\partial v_{\nu}}{\partial \tau_{\nu}} - u_{\nu} + 
    S_{\nu}\right),\\
\label{eq:FeautRTEv}
\frac{\partial v_{\nu}}{\partial t} & = c\chi_{\nu}\left(
    \frac{\partial u_{\nu}}{\partial \tau_{\nu}} - v_{\nu}\right),
\end{align}
where $S_{\nu} = \eta_{\nu} / \chi_{\nu}$ is the source 
function.
Similarly, the boundary conditions can be rewritten as:
\begin{gather}
\label{eq:FeautBC1}
u_{\nu}(t, z_{\rm o}, \mu) = v_{\nu}(t, z_{\rm o}, \mu) + 
                               I_{\nu}^{\rm ext}(t, \mu),\\
\label{eq:FeautBC2}
u_{\nu}(t, z_{\rm min}, \mu) = -v_{\nu}(t, z_{\rm min}, \mu).
\end{gather}
Equations~\eqref{eq:FeautRTEu} and \eqref{eq:FeautRTEv}, along with 
the boundary conditions \eqref{eq:FeautBC1} and \eqref{eq:FeautBC2}, 
form the basis of our numerical solution to the RTE.

\subsection{Atomic Rate Equations}
\label{subsect:AtomicRates}

The extinction and emission coefficients for the radiation field 
depend directly on the level populations 
of the various bound atomic states, as well as the degree of ionization 
of the medium. 
The rate equation for each level $j$ can be written:
\begin{equation}
\label{eq:AtomicRates}
\frac{d n_j}{dt} = \sum_{l} \Bigl[\left(C_{lj} + R_{lj}\right) n_l - 
                                   \left(C_{jl} + R_{jl}\right) n_j\Bigr],
\end{equation}
where $R_{lj}$ and $C_{lj}$ are the rate coefficients for radiative and 
collisional transitions between levels $l\rightarrow j$, respectively.  
The sum $\Sigma_l$ is over all atomic levels $l\ne j$.
Using the notation of \S\ref{subsect:RTE}, we write the radiative rates as:
\begin{equation}
\label{eq:RateCoef}
\begin{split}
R_{lj} & \equiv \int_{4\pi} d\Omega 
              \int_0^{\infty} \frac{d\nu}{h\nu} \left(U_{lj} + 
                                              V_{lj} I_{\nu}\right)\\
       & = 4\pi \int_0^1 d\mu \int_0^{\infty} \frac{d\nu}{h\nu} 
         \left(U_{lj} + V_{lj} u_{\nu}\right),
\end{split}
\end{equation}
where in the second equality we assume that the line profile for the 
transition is symmetric in $\mu$.  The collisional rate coefficients 
are assumed to have a dependence on the electron number density, 
$n_e$, and temperature, $T_e$.  For the purposes of our calculations, 
we assume that $n_e$ and $T_e$ are known functions of $t$ and $z$, 
or that they may be calculated separately from the atomic level 
populations, using a previous iteration in our numerical solution 
(see \S\ref{sect:Numerical}).

Equation~\eqref{eq:AtomicRates} must be supplemented by an initial 
condition for each level $j$.  These are determined by solving the 
time independent atomic rate equations, obtained from 
equation~\eqref{eq:AtomicRates} by setting 
$d n_j / dt \rightarrow 0$, and using $u_{\nu}(t=0, z, \mu)$ for 
the radiation field (see \S\ref{subsect:RTE}).  Note that the 
steady-state calculations determining our initial conditions are 
equivalent to the problem described by \citet[][]{RybickiHummer92a}, 
with an external source added to the boundary conditions.

The primary difficulty in computing solutions for 
the radiation field and level populations are that 
equations~\eqref{eq:FeautRTEu}, \eqref{eq:FeautRTEv}, and 
\eqref{eq:AtomicRates} are coupled.  Thus, the populations, 
$n_j$, must be determined simultaneously with the radiation 
field, $u_{\nu}$.  In \S\ref{sect:Numerical} we describe a numerical 
method for calculating self consistent values for these 
variables.

\section{Numerical Method}
\label{sect:Numerical}

\subsection{Discretization Scheme}
\label{subsect:Discretization}

We solve the RTE and atomic rate equations using 
numerical quadrature on discrete grids.  
For the reference optical depth, we define a set of $D$ points, 
$\{\tau\}_{d=0}^{D-1}$, that are equally spaced logarithmically 
over the interval $[\tau_{\rm min}, \tau_{\rm max}]$.  The number of 
spatial points 
$D$ is chosen such that there is adequate resolution over the full range 
of optical depth for photons in the atomic transitions.
In the models of \S\ref{sect:Results}, we set 
$\tau_{\rm min} = 10^{-4}$ and $\tau_{\rm max} = 10^3$.

For the time grid, we define a set of $K$ points, $\{t_k\}_{k=0}^{K-1}$.  
The appropriate spacing and interval for the time grid depends on the 
model used for the external radiation, as well as the values of the parameters 
in equation~\eqref{eq:lightcurve}.  
As discussed in \S\ref{sect:Introduction}, 
we are primarily interested in the intermediate regime for which 
$T_{\rm VR} \gtrsim T_{\rm ref}$ and $T_{\rm VR}\lesssim T_{\rm LC}$; 
we therefore choose values of the parameters that reproduce this 
behaviour (see \S\ref{sect:Results}).  When using
equation \eqref{eq:lightcurve} for the external radiation, 
we employ a linearly spaced time grid 
spanning an interval equal to several light crossing times.  
When using the 
ramp model in our comparisons to previous work, we employ a 
logarithmically spaced time grid that exceeds the light crossing time by 
more than an order of magnitude.  
The use of linear spacing for the time grid in the former 
case is important, as resolution 
at the smallest time variation scale is needed for each decade of time 
in the integration interval.  

For the frequency grid, we use the variable
\begin{gather}
x \equiv (\nu - \nu_{lj}) / \Delta\nu_{lj},\\
\Delta\nu_{lj} \equiv \nu_{lj} \left(b / c\right),
\end{gather}
where $x$ is the number of Doppler widths, $\Delta\nu_{lj}$, from line centre 
of the transition $l\rightarrow j$.  In \S\ref{sect:Results}, we present 
results of calculations using Doppler and Voigt line profiles.  For the 
Doppler case, we define a set of $N$ points, 
$\{x\}_{n = 0}^{N-1}$, spaced linearly in the interval $[0, 4]$.  For the 
Voigt case with parameter $a = 0.01$, we augment the Doppler grid
with a set of logarithmically spaced points out to 
$x = 20$.  We assume that $\Delta\nu_{lj}$ and $a$ are the same for all 
transitions in our atomic model; we can therefore reuse the same frequency 
grid for each transition.  It is straightforward to define a set of 
frequency integration weights, $h_n$, for numerical quadrature.  In the 
calculations that follow, we use the extended trapezoidal rule to define 
the integration weights, breaking the integral into linear Doppler and 
logarithmic Voigt sections as necessary.

It is standard to define a set of $M$ angular grid points, 
$\{\mu_m\}_{m=0}^{M-1}$, 
such that each $\mu_m$ is an abscissa for a Gaussian quadrature rule with 
corresponding integration weight $w_m$ 
\citep[see, e.g.,][]{Mihalas78a, Cannon85a}.
It is often convenient, based on the form of the angular integrals, to 
choose the weights and abscissas for Gauss-Legendre quadrature.  For an 
isotropic external source, this choice is adequate.  
However, for a highly collimated source, 
the angular variation is such that 
the external radiation is non-zero for a narrow range of 
angles centred on $\theta = \pi$; this range  
is much smaller than the grid resolution of our models.
In this case, we have found it useful to use Gauss-Lobatto quadrature, 
which includes the endpoint of the integration interval, $\mu = 1$, 
corresponding to radiation traveling in the $\theta = \pi$ 
direction for the radiation field $I_{\nu}^-$.  
Recall that we restrict $\mu\in [0, 1]$, and have defined 
$I_{\nu}^{\rm ext}(t, \mu) = I_{\nu}^-(t, z_{\rm o}, \mu)$ to consist 
of external photons traveling in directions with $\theta > \pi / 2$.
The value of the angle-averaged external intensity is fixed at 
$J_{\nu}^{\rm ext}(t) \equiv \frac{1}{2}\int_0^{1} 
d\mu I_{\nu}^{\rm ext}(t, \mu)$. 
We then set $w_0 I_{\nu}^{\rm ext}(t, \mu_0) / 2 = J_{\nu}^{\rm ext}(t)$, 
where $w_0$ and $\mu_0$ are the integration weight and abscissa associated 
with $\mu = 1$ ($\theta = \pi$).  
These are the only non-zero values of $I_{\nu}^{\rm ext}$ in 
the boundary condition, and can be used directly with 
equation~\eqref{eq:FeautBC1}. 
This procedure is equivalent to solving two sets of coupled equations for the 
collimated and 
angularly redistributed components of the radiation field.

It is convenient to use a single index that contains the combined frequency 
and angle dependences of the radiation variables.  We define 
$Q = N\times M$ grid points with the formula $i\equiv n + m N$.  The 
values of a function, $g$, on the grids defined above, are written 
$g_{\nu_n}(t_k, \tau_d, \mu_m) \equiv g_{kdi}$.  We will use this notation 
extensively in the work that follows. 

\subsection{Solution to the Radiative Transfer Equation}
\label{subsect:RTESol}

To solve the partial differential equations (PDEs) in 
\eqref{eq:FeautRTEu} and \eqref{eq:FeautRTEv}, we follow 
\citet[][]{Kunasz83a} and use a method of lines approach, in 
which we replace the PDEs with ordinary differential equations (ODEs) by 
discretizing the spatial derivatives.  It will prove useful to 
augment our spatial grid by adding a set of $D - 1$ points, 
$\{\tau_{d+1/2}\}_{d=0}^{D-2}$,
that are located between and equidistant from 
$\tau_d$ and $\tau_{d+1}$.  If equations~\eqref{eq:FeautRTEu} 
and \eqref{eq:FeautRTEv} are evaluated at 
$\tau_d$ and $\tau_{d+1/2}$, respectively, we obtain:
\begin{align}
\label{eq:uRTE_d}
\frac{d u_{di}}{d t} & \equiv f^u_{di} =\nonumber\\ 
    & c\chi_{di}\left(
    \frac{v_{d+1/2,i} - v_{d-1/2,i}}{\Delta \tau_{di}} - u_{di} + 
    S_{di}\right),\\
\label{eq:vRTE_dph}
\frac{d v_{d+1/2,i}}{d t} & \equiv f^v_{d+1/2,i} =\nonumber\\
    & c\chi_{d+1/2,i}\left(
    \frac{u_{d+1,i} - u_{di}}{\Delta\tau_{di}^+} - v_{d+1/2,i}\right).
\end{align}
We use the following notation:
\begin{align}
\label{eq:Dtau_p}
\Delta\tau_{di}^+ & \equiv \tau_{d+1,i} - \tau_{di},\\
\label{eq:Dtau_m}
\Delta\tau_{di}^- & \equiv \tau_{di} - \tau_{d-1,i},\\
\label{eq:Dtau}
\Delta\tau_{di} & \equiv \tau_{d+1/2,i} - \tau_{d-1/2,i}\nonumber\\
    & = \left(\Delta\tau_{di}^+ + \Delta\tau_{di}^-\right) / 2.
\end{align}
Equations~\eqref{eq:uRTE_d} and \eqref{eq:vRTE_dph} represent a set of 
$(2D - 4)Q$ coupled ODEs for $d = 1, \ldots, D - 2$ and $i = 0, \ldots Q-1$.
Equations for the variables at the boundaries will be considered below.

The time derivatives can be integrated according to the 
`$\theta$ method', which parametrizes the solution in 
terms of the Backward Euler and Crank-Nicholson solutions:
\begin{gather}
\label{eq:uRTE_sol}
u_{kdi} = u_{k-1,di} + \Delta t_k\left(\bar{\theta} f^u_{kdi} + 
   \bar{\eta} f^u_{k-1,di}\right),\\
\label{eq:vRTE_sol}
\begin{split}
v_{k,d+1/2,i} & = v_{k-1,d+1/2,i} +\\ 
   & \Delta t_k\left(\bar{\theta} f^v_{k,d+1/2,i} +
   \bar{\eta} f^v_{k-1,d+1/2,i}\right),
\end{split}
\end{gather}
where $\Delta t_k \equiv t_k - t_{k-1}$ and 
$\bar{\eta}\equiv 1 - \bar{\theta}$.
The parameter $\bar{\theta}$ is defined such that 
equations~\eqref{eq:uRTE_sol} and \eqref{eq:vRTE_sol} are integrated 
by the Backward Euler method for $\bar{\theta} = 1$, and by the 
Crank-Nicholson method for $\bar{\theta} = 1/2$.  We explore more 
complicated Runge-Kutta integration schemes in Appendix~\ref{sect:RKMethods}.

Equation~\eqref{eq:vRTE_sol} can be used to eliminate the variables 
$v_{k,d\pm1/2,i}$ in \eqref{eq:uRTE_sol}.  Substituting \eqref{eq:vRTE_sol} 
into \eqref{eq:uRTE_sol} and using the explicit forms for $f^{u,v}_{kdi}$ in 
\eqref{eq:uRTE_d}--\eqref{eq:vRTE_dph} yields the following system 
of equations:\footnote{Our expressions 
differ slightly from 
the corresponding equations (9a)--(10) in \citet{Kunasz83a}.  
We can recover 
the equations in the earlier work by setting $\bar{\eta}_{kdi}, 
\bar{\eta}_{k,d\pm 1/2,i}\rightarrow \bar{\eta}$ and substituting the explicit 
form for the two-level source function in our formulas.  Numerical tests 
showed no 
significant difference in the solutions using either form for the two-level 
atom.}
\begin{equation}
\label{eq:DiscRTE}
\begin{split}
-A_{kdi} u_{k,d-1,i} + B_{kdi} u_{kdi} - C_{kdi} u_{k,d+1,i}\\ = 
    \bar{\theta} S_{kdi}
    + \bar{\eta}_{kdi} S_{k-1,di} + F_{kdi} u_{k-1,d-1,i}\\ + G_{kdi} u_{k-1,di}
    + H_{kdi} u_{k-1,d+1,i}\\ + T^+_{kdi} v_{k-1,d+1/2,i}
    - T^-_{kdi} v_{k-1,d-1/2,i},
\end{split}
\end{equation}
where
\begin{align}
A_{kdi} & \equiv \frac{\bar{\theta}^2}{\left(\delta_{k,d-1/2,i} + \bar{\theta}\right) 
                                 \Delta\tau_{kdi}^-\Delta\tau_{kdi}},\\
C_{kdi} & \equiv \frac{\bar{\theta}^2}{\left(\delta_{k,d+1/2,i} + \bar{\theta}\right)
                                 \Delta\tau_{kdi}^+\Delta\tau_{kdi}},\\
B_{kdi} & \equiv \delta_{kdi} + \bar{\theta} + A_{kdi} + C_{kdi},
\end{align}
and
\begin{align}
F_{kdi} & \equiv \frac{\bar{\theta}\bar{\eta}_{k,d-1/2,i}}{\left(\delta_{k,d-1/2,i} + 
   \bar{\theta}\right)
   \Delta\tau_{k-1,di}^-\Delta\tau_{kdi}},\\
H_{kdi} & \equiv \frac{\bar{\theta}\bar{\eta}_{k,d+1/2,i}}{\left(\delta_{k,d+1/2,i} + 
   \bar{\theta}\right)
   \Delta\tau_{k-1,di}^+\Delta\tau_{kdi}},\\
G_{kdi} & \equiv \delta_{kdi} - \bar{\eta}_{kdi} - F_{kdi} - G_{kdi},\\
T^{\pm}_{kdi} & \equiv \frac{\bar{\eta}_{kdi}}{\Delta\tau_{k-1,di}} + 
   \frac{\bar{\theta}}{\Delta\tau_{kdi}}
   \frac{\delta_{k,d\pm1/2,i} - \bar{\eta}_{k,d\pm 1/2,i}}
   {\delta_{k,d\pm1/2,i} + \bar{\theta}},\\
\end{align}
using the definitions
\begin{align}
\delta_{kdi} & \equiv 1 / \left(c\chi_{kdi}\Delta t_k\right),\\
\delta_{k,d\pm 1/2,i} & \equiv 1 / \left(c\chi_{k,d\pm 1/2,i}\Delta t_k\right),\\
\bar{\eta}_{kdi} & \equiv \bar{\eta}\left(\chi_{k-1,di} / \chi_{kdi}\right),\\
\bar{\eta}_{k,d\pm 1/2,i} & \equiv 
   \bar{\eta}\left(\chi_{k-1,d\pm 1/2,i} / \chi_{k,d\pm 1/2,i}\right).
\end{align}

Equation~\eqref{eq:DiscRTE} represents a set of $D-2$ equations for each value 
of $k$ and $i$.  To complete the system of equations, we spatially discretize 
\eqref{eq:FeautRTEv} to first order at $d = 0$ and $d = D - 1$, and use the 
boundary conditions \eqref{eq:FeautBC1}--\eqref{eq:FeautBC2} to eliminate 
the variables $v_{k,d=0,i}$ and $v_{k,d=D-1,i}$.  If the resulting expressions 
are integrated with respect to time as described above, we obtain:
\begin{equation}
\label{eq:DiscBC1}
\begin{split}
\left(\delta_{k,d=0,i} + \bar{\theta} + 
C_{k,d=0,i}\right) u_{k,d=0,i} -
C_{k,d=0,i} u_{k,d=1,i}\\ = 
\left(\delta_{k,d=0,i} - \bar{\eta}_{k,d=0,i} - 
H_{k,d=0,i}\right) u_{k-1,d=0,i}\\ -
H_{k,d=0,i} u_{k-1,d=1,i}
+ \left(\delta_{k,d=0,i} + \bar{\theta}\right) I^{\rm ext}_{ki}\\
+ \left(\bar{\eta}_{k,d=0,i} - \delta_{k,d=0,i}\right) I^{\rm ext}_{k-1,i},
\end{split}
\end{equation}
and
\begin{equation}
\label{eq:DiscBC2}
\begin{split}
-A_{k,d=D-1,i} & u_{k,D-2,i}\\
+ \bigl(\delta_{k,d=D-1,i} + \bar{\theta} 
+ & A_{k,d=D-1,i}\bigr) u_{k,d=D-1,i}\\
= & G_{k,d=D-1,i} u_{k-1,D-2,i}\\
+ \bigl(\delta_{k,D-1,i} - \bar{\eta}_{k,D-1,i} - 
      & G_{k,d=D-1,i}\bigr) u_{k-1,D-1,i},
\end{split}
\end{equation}
where
\begin{align}
C_{k,d=0,i} & \equiv \frac{\bar{\theta}}{\Delta\tau_{k,d=0,i}^+},\\
H_{k,d=0,i} & \equiv \frac{\bar{\eta}_{k,d=0,i}}{\Delta\tau_{k-1,d=0,i}^+},\\
A_{k,d=D-1,i} & \equiv \frac{\bar{\theta}}{\Delta\tau_{k,D-1,i}^-},\\
G_{k,d=D-1,i} & \equiv \frac{\bar{\eta}_{k,D-1,i}}{\Delta\tau_{k-1,D-1,i}^-}.
\end{align}
Thus, if the radiation field is known for all values of $d$ and $i$ at $k - 1$, 
then equations~\eqref{eq:DiscRTE}, \eqref{eq:DiscBC1}, and \eqref{eq:DiscBC2} 
represent $Q$ tridiagonal, linear systems for the spatial variation of the 
radiation 
variables, $u_{kdi}$, at time $t_k$.  In matrix form, the systems of equations 
can be written:
\begin{equation}
\label{eq:MatrixRTE}
{\bf L}_{ki} \cdot {\bf u}_{ki} = \bar{\theta}{\bf S}_{ki} + 
	\bar{\bf S}_{k-1,i} + 
	{\bf B}_{ki},
\end{equation}
where ${\bf u}_{ki}$, ${\bf S}_{ki}$, and 
$\bar{\bf S}_{k-1,i}$ are $D\times 1$ column vectors with 
components $u_{kdi}$, $S_{kdi}$, and $\bar{\eta}_{kdi} S_{k-1,di}$, 
respectively.  The symbol ${\bf L}_{ki}$ 
denotes a tridiagonal matrix, while the column vector ${\bf B}_{ki}$ depends 
on the values of the variables at the previous time step, as well as the 
boundary conditions;
the components of ${\bf L}_{ki}$ and ${\bf B}_{ki}$ are directly determined 
by equations~\eqref{eq:DiscRTE}, \eqref{eq:DiscBC1}, and \eqref{eq:DiscBC2}.

The initial condition for the radiation field, $u_{k=0,di}$, is determined 
by the prescription of \S\ref{subsect:RTE}: the time independent 
RTE is solved using the value of the source function at $t_{k=0}$.  The 
equations governing the time independent equation are well known 
(see, e.g., Chapter 6 of \citealt[][]{Mihalas78a}, or Appendix~A 
of \citealt[][]{RybickiHummer91a}).  The formal solution to the TDRTE
can then be obtained from equation~\eqref{eq:MatrixRTE} and the initial 
condition by solving the tridiagonal system for successive 
$k = 1, \ldots K - 1$.  The results can be structured as a block matrix 
equation:
\begin{equation}
\label{eq:FormalRTE}
{\bf\tilde{u}}_i = 
\tilde{\boldsymbol{\Lambda}}_{i} \cdot 
\tilde{\bf S}_i
+ \tilde{{\bf B}}_i,
\end{equation}
where $\tilde{\boldsymbol{\Lambda}}_{i}$ is a $K\times K$ block 
lower-triangular matrix, 
the elements of which are $D\times D$ matrices, and 
$\tilde{\bf u}_i$, $\tilde{\bf S}_i$, $\tilde{\bf B}_i$ are $K\times 1$ 
block vectors, the components 
of which are $D\times 1$ column vectors.  
The block vector $\tilde{\bf B}_i$ has a complicated dependence on the 
initial and boundary conditions for the solution.
The structure of the formal solution in terms of the 
source function values, $S_{kdi}$, is also complicated, with each pair of 
$(k, d)$ indices coupling to other spatial grid points at previous times.

A key property of equation~\eqref{eq:MatrixRTE} is that, when  
$u_{k-1,di}$ and $v_{k-1,d\pm 1/2,i}$ are known, it can be solved as a 
tridiagonal 
matrix equation of dimension $D$, independently for each frequency-angle 
index $i$.
Using the linear algebraic methods described in Appendices~A and B of 
\citet[][]{RybickiHummer91a}, the values of $u_{kdi}$ (and hence the formal 
solution) can be computed in $\mathcal{O}(KDQ)$ operations.  In addition, 
the diagonal elements of ${\bf L}_{ki}^{-1}$, which will be used for the 
simultaneous solution to the RTE and rate equations in \S\ref{subsect:ALI}, 
can be calculated simultaneously at little additional computational cost.

\subsection{Solution to Atomic Rate Equations}
\label{subsect:RatesSol}

The equations governing the atomic level populations in 
\eqref{eq:AtomicRates} 
form a system of ODEs.  Though the system exhibits no explicit coupling of the 
populations at different spatial grid points, coupling is 
introduced implicitly by the presence of the radiation field in the
rate coefficients, $R_{lj}$.

The time derivative in equation~\eqref{eq:AtomicRates} can be integrated with
the same technique used for the RTEs in \S\ref{subsect:RTESol}.  Applying the 
$\theta$~method, we obtain:
\begin{gather}
\label{eq:DiscRates}
n_{j,kd} = n_{j,k-1,d} + \Delta t_k \left(\bar{\theta} f_{j,kd} + 
                                          \bar{\eta} f_{j,k-1,d}\right),\\
\label{eq:RatesRHS}
\begin{split}
f_{j,kd} \equiv \sum_l \Bigl[\left(C_{lj,kd} + R_{lj,kd}\right) n_{l,kd} 
   \\- \left(C_{jl,kd} + R_{jl,kd}\right) n_{j,kd}\Bigr].
\end{split}
\end{gather}
The radiative rate coefficients take the form:
\begin{equation}
\label{eq:Rrates}
R_{lj,kd} = \frac{4\pi}{h} \sum_{i=0}^{Q-1} \frac{q_i}{\nu_n} 
    \left[U_{lj,i} + V_{lj,i}\, u_{kdi}\right],
\end{equation}
where the integration weights for the double integral in 
equation~\eqref{eq:RateCoef} are given by $q_i = h_n w_m$.
From equation~\eqref{eq:FormalRTE}, the expression for 
$u_{kdi}$ contains terms of the form 
$\bigl[\tilde{\boldsymbol{\Lambda}}_i\cdot \tilde{\bf S}_i\bigr]_{kd}$, 
which couple various $S_{kdi}$ for different values of $k$ and $d$.  
Because $S_{kdi}$ has a non-linear dependence on the level populations 
$n_j$, equations~\eqref{eq:DiscRates} and \eqref{eq:RatesRHS} represent a 
non-linear system of equations for the populations, coupled in the 
indices $j$, $k$, and $d$.

The initial conditions for the $n_j$ are obtained from 
equation~\eqref{eq:AtomicRates} by setting the derivative on the 
left-hand side equal to zero, and using the initial condition for 
the radiation field, $u_{k=0,di}$ in the rate coefficients 
(see \S\ref{subsect:RTESol}).  This 
results in a coupled non-linear system in the indices $j$ and $d$.

\subsection{Simultaneous Solution with Accelerated Lambda Iteration}
\label{subsect:ALI}

The simplest simultaneous solution method for the RTE and 
atomic rate equations, often referred to as 
`Lambda Iteration', can be outlined as follows:
\begin{enumerate}
\item{Start with a set of values 
for the level populations at all grid points, 
denoted $n_{j,kd}^{\dag}$.  These populations determine the 
values $\chi_{kdi}^{\dag}$ and $\eta_{kdi}^{\dag}$ in 
equations~\eqref{eq:chi} and \eqref{eq:eta}.}
\item{Using these values, compute the formal solution for the radiation field, 
$u_{kdi}^{\dag}$ (where the symbol $\dag$ indicates 
that the populations $n_{j,kd}^{\dag}$ have been used to obtain the 
field).}
\item{Substitute the field $u_{kdi}^{\dag}$ into the linear systems of 
equation~\eqref{eq:DiscRates} and solve for an updated set of populations 
$n_{j,kd}$.}
\item{Check for convergence of the populations by forming the 
quantity:
\begin{equation}
\Theta \equiv \text{\it Max}\left(\frac{\left|n_{j,kd} - 
                                               n_{j,kd}^{\dag}\right|}
                                   {\epsilon_A + \epsilon_R n_{j,kd}}\right),
\end{equation}
where $\epsilon_A$ and $\epsilon_R$ are the absolute and relative local error 
tolerances, respectively (chosen for the particular problem to be solved).  
The symbol {\it Max}\, indicates that the largest value for the quantity in 
parentheses, considered for each combination of $j, k,$ and $d$, should be 
assigned to $\Theta$.}
\item{If $\Theta \le 1$, we consider the system of equations to have converged, 
representing a self consistent solution for the field and level 
populations.  If $\Theta > 1$, we replace 
$n_{j,kd} \rightarrow n_{j,kd}^{\dag}$, and repeat steps $1-4$.}
\end{enumerate}

Unfortunately, this scheme suffers from several deficiencies, which render it 
unusable for many practical problems with $\tau_{\rm max} \gg 1$.  
For detailed discussions of these issues, 
see, for example, \citet{Mihalas78a}, \citet{Auer91a}, and the 
references therein.

We can improve on the Lambda iteration scheme by using the ALI method.  This 
technique replaces the formal solution of equation~\eqref{eq:FormalRTE} with 
an approximate expression to be used in the iteration scheme:
\begin{equation}
\label{eq:FormalApprox}
\begin{split}
{\bf\tilde{u}}_i & = {\boldsymbol{\tilde{\Lambda}}}_i^* \cdot {\bf\tilde{S}}_i 
                   + \left({\boldsymbol{\tilde{\Lambda}}}_i - 
                           {\boldsymbol{\tilde{\Lambda}}}_i^*\right) \cdot
                   {\bf\tilde{S}}_i^{\dag} + 
		   {\tilde{\bf B}}_i^{\dag} \\ & =
                   {\boldsymbol{\tilde{\Lambda}}}_i^* \cdot 
                   \left({\bf\tilde{S}}_i - {\bf\tilde{S}}_i^{\dag}\right) 
                   + {\bf\tilde{u}}_i^{\dag},
\end{split}
\end{equation}
where ${\boldsymbol{\tilde{\Lambda}}}_i^*$ is an appropriately chosen 
approximation to the full matrix, and the symbol $\dag$ indicates that 
${\bf\tilde{S}}_i^{\dag}$, $\tilde{\bf B}_i^{\dag}$, and 
${\bf\tilde{u}}_i^{\dag}$ have been calculated using 
the quantities $n_{j,kd}^{\dag}$.  As the level populations converge, 
$n_{j,kd}^{\dag}\rightarrow n_{j,kd}$, and 
equation~\eqref{eq:FormalApprox} becomes 
identical to equation~\eqref{eq:FormalRTE}. 
 
As originally shown by \citet[][]{Olsonetal86a}, an efficient choice 
for $\tilde{\boldsymbol{\Lambda}}^*_i$ is the diagonal of the full matrix 
$\tilde{\boldsymbol{\Lambda}}_i$.  With this choice, we can write 
equation~\eqref{eq:FormalApprox} as:
\begin{equation}
\label{eq:FormalSimp}
u_{kdi} = \bar{\theta} \Lambda_{kdi}^* \left(S_{kdi} - S_{kdi}^{\dag}\right) + 
          u_{kdi}^{\dag},
\end{equation}
where $\Lambda_{kdi}^*\equiv \left[{\bf L}_{ki}^{-1}\right]_{dd}$ denotes the 
diagonal elements of the matrix $\tilde{\boldsymbol{\Lambda}}_i$, which are 
calculated as described in \S\ref{subsect:RTESol}.

In principle, more complicated choices for 
$\tilde{\boldsymbol{\Lambda}}^*$ can lead to faster convergence.  For example, 
instead of using the diagonal of the full matrix, we could use the tridiagonal 
submatrix of $\tilde{\boldsymbol{\Lambda}}^*$.  This leads to a more 
complicated expression than \eqref{eq:FormalSimp}, involving 
elements that cannot be obtained easily from ${\bf L}_{ki}^{-1}$;
thus, considerable computational effort is required to use the ALI method 
in this case.  This situation is well known for multidimensional radiative 
transfer problems \citep[see, e.g.,][]{KunaszOlson88a, Aueretal94a}.

To implement the ALI method, we follow \citet[][]{RybickiHummer92a} and 
use an alternate formulation of equation~\eqref{eq:FormalSimp}:
\begin{equation}
\label{eq:PsiApprox}
\begin{split}
u_{kdi} & = \bar{\theta} \Psi_{kdi}^* \left(\eta_{kdi} - 
	\eta_{kdi}^{\dag}\right) + u_{kdi}^{\dag}\\
	& = \bar{\theta} \Psi_{kdi}^* \sum_{l, j} U_{lj,i} \left(
               n_{l,kd} - n_{l,kd}^{\dag}\right) + u_{kdi}^{\dag}.
\end{split}
\end{equation}
Unlike the line transitions, the background and scattering contributions 
to $\eta$ are 
treated as quantities with values that are either fixed in the 
problem under consideration, or that can be approximated using the 
results from a previous iteration.  Therefore, the quantities $\eta_c$ cancel 
in equation~\eqref{eq:PsiApprox} 
\citep[c.f., the discussion in \S2.1 of][]{RybickiHummer92a}.

The new matrix $\Psi^*_{kdi}$ is related to 
$\Lambda^*_{kdi}$ through 
\begin{equation}
\label{eq:Psi}
\Psi_{kdi}^* = \Lambda^*_{kdi} / \chi_{kdi}^{\dag}.  
\end{equation}
The $\dag$ symbol appears in this equation because the 
approximate matrices are themselves constructed using the quantities 
$n_{j,kd}^{\dag}$.  The advantage of using 
$\Psi^*_{kdi}$ is 
that $u_{kdi}$ now depends linearly on the updated quantities 
$n_{j,kd}$ through $\eta_{kdi}$, whereas $S_{kdi}$ exhibits a non-linear 
dependence due to the factor of $\chi_{kdi}$ in the denominator.

The combination of equations~\eqref{eq:Rrates}, and \eqref{eq:PsiApprox} 
yields:
\begin{equation}
\label{eq:ALIRHS}
\begin{split}
R_{lj,kd} & n_{l,kd} - R_{jl,kd} n_{j,kd} \\
        = & \frac{4\pi}{h} \sum_{i=0}^{Q-1} \frac{q_i}{\nu_n} \Biggl\{
	U_{lj,i} n_{l,kd} - U_{jl,i} n_{j,kd}
	+ \chi_{lj,kdi}\\
	\times & \Biggl[
	\bar{\theta} \Psi_{kdi}^* \sum_{l',j'} U_{l'j', i} \left(n_{l',kd} - 
	n_{l',kd}^{\dag}\right) + u_{kdi}^{\dag}
	\Biggr]\Biggr\}.
\end{split}
\end{equation}
Non-linearities occur in equation~\eqref{eq:ALIRHS} due to terms of the 
form $\chi_{lj,kdi} n_{l',kd}$.  
Following 
\citet{RybickiHummer91a, RybickiHummer92a}, we linearize the equation 
by making the replacement $\chi_{lj,kdi} n_{l',kd} \rightarrow 
\chi_{lj,kdi}^{\dag} n_{l',kd}$.\footnote{As noted by 
\citet{RybickiHummer91a}, the 
alternative procedure $\chi_{lj,kdi} n_{l',kd} \rightarrow 
\chi_{lj,kdi} n_{l',kd}^{\dag}$ recovers the Lambda Iteration scheme.}  
This procedure results in the `preconditioned' expressions:
\begin{equation}
\label{eq:ALIRHSLIN}
\begin{split}
R_{lj,kd} & n_{l,kd} - R_{jl,kd} n_{j,kd} \\
        = & \frac{4\pi}{h} \sum_{i=0}^{Q-1} \frac{q_i}{\nu_n} \Biggl[
	U_{lj,i} n_{l,kd} - U_{jl,i} n_{j,kd}
	+ \bar{\theta} \Psi_{kdi}^*\\
	& \times \sum_{l',j'} U_{l'j', i} 
	\bigl(\chi_{lj,kdi}^{\dag} 
	n_{l',kd} - \chi_{lj,kdi} n_{l',kd}^{\dag}\bigr)\\
	& + \chi_{lj,kdi} u_{kdi}^{\dag}\Biggr].
\end{split}
\end{equation}
We can rewrite equation~\eqref{eq:ALIRHSLIN} using our RTE formal solution and 
rate equation discretization.  Employing identity~\eqref{eq:Psi} and collecting 
terms about $n_{l,kd}$, $n_{j,kd}$, and $n_{l',kd}$, we obtain:
\begin{equation}
\label{eq:ALIRHSFIN}
\begin{split}
f_{j,kd} = \sum_l\Bigl[& \bigl(C_{lj,kd} + \mathcal{R}_{lj,kd}\bigr) n_{l,kd}\\
		     - & \bigl(C_{jl,kd} + \mathcal{R}_{jl,kd}\bigr)n_{j,kd}\\
		     + & \sum_{l'\ne l,j} 
		       \mathcal{R}_{lj,kd}^{l'} n_{l',kd}\Bigr],
\end{split}
\end{equation}
where
\begin{equation}
\begin{split}
\label{eq:R_lj}
\mathcal{R}_{lj,kd} \equiv \frac{4\pi}{h} \sum_{i=0}^{Q-1} \frac{q_i}{\nu_n}
   \left\{
   U_{lj,i}\left[1 - \left(\chi_{jl,kdi}^{\dag} / \chi_{kdi}^{\dag}\right)
   \bar{\theta} \Lambda_{kdi}^*\right]\right.\\\left.
   + V_{lj,i}\left[u_{kdi}^{\dag} 
   - \bar{\theta} \Lambda_{kdi}^* 
   \left(S_{kdi}^{\dag} - 
   \eta_{c,kdi}^{\dag} / \chi_{kdi}^{\dag}\right)\right]\right\}
\end{split}
\end{equation}
and
\begin{equation}
\label{eq:R_lj^l'}
\mathcal{R}_{lj,kd}^{l'} \equiv \frac{4\pi}{h} \sum_{i=0}^{Q-1} 
   \frac{q_i}{\nu_n} \left(\chi_{lj,kdi}^{\dag} / \chi_{kdi}^{\dag}\right)
   \bar{\theta} \Lambda_{kdi}^* \sum_{j'} U_{l'j',i}.
\end{equation}
Substitution of equations~\eqref{eq:ALIRHSFIN}--\eqref{eq:R_lj^l'} 
into \eqref{eq:DiscRates} yields explicit expressions for the 
linear systems.

Lambda Iteration can now be replaced by an improved ALI scheme that exhibits 
faster convergence.  The algorithm is essentially the same as that for 
Lambda Iteration except that the modified 
linear system is used to compute the updated level populations. 

The ALI rate equations have some attractive properties.  
Given the initial 
conditions $n_{j,k=0,d}$, $u_{k=0,di}$, and taking $\Lambda_{k=0,di}^* = 0$, 
the ALI iteration 
can be applied to successive $k = 1, \ldots K-1$, 
using the values 
$\Lambda_{k-1,di}^*$, $\chi_{k-1,di}^{\dag}$, and $S_{k-1,di}^{\dag}$ from 
the previous time step to initialize the ALI iteration in the next time 
step.  
If $L$ is the number 
of levels in the atomic model, the solution of the rate equations
at a given time step requires $\mathcal{O}(DL^3)$ operations,  
and each iteration requires $\mathcal{O}(DQL^3)$ operations.
Therefore, in the numerical implementation of our solution, the 
time steps form
an outer loop, while the iteration over the level populations forms an 
inner loop.  
This design for our iterative solution is analogous to that used by 
\citet[][]{HubenyBurrows07a}, who noted that 
employing level populations from the previous step to seed the 
iteration of the current step significantly increases 
the convergence rate in a time dependent calculation.  
Because the structure of our solution method is the same as in
their work, our calculations 
can be straightforwardly implemented in applications requiring 
radiation hydrodynamics.

It should be noted that one unattractive property of the ALI equations
is that they introduce coupling between atomic levels that are not associated 
by transitions in the original rate equations, due to the terms containing  
$R_{lj,kd}^{l'}$.
While equation~\eqref{eq:ALIRHSFIN} can be implemented directly with 
\eqref{eq:DiscRates}, 
for many problems of interest substantial simplifications can be achieved.  
As discussed in \S\S2.4-2.5 of \citet{RybickiHummer92a}, for atomic systems 
containing transitions that don't exhibit `significant' frequency overlap, 
terms of the form $\chi_{lj,kdi} U_{l'j',i}$ can be neglected when $lj$ 
and $l'j'$ denote different transitions,
by setting $R_{lj,kd}^{l'}\rightarrow 0$ in 
equation~\eqref{eq:ALIRHSFIN}.
The resulting linear system only couples levels that are associated by 
transitions in the original rate equations.

In the following, we exclusively consider systems that exhibit negligible
frequency overlap in their transitions.  For fixed $lj$, only a 
single transition contributes to the extinction and emission, which are 
non-zero for a subset of the total frequency grid.
In this range, $\chi_{lj,kdi}^{\dag}\rightarrow \chi_{kdi}^{\dag}$, 
and the terms in \eqref{eq:R_lj} 
become\footnote{We note that the result \eqref{eq:ALISimp} can be obtained 
directly by 
substituting equation~\eqref{eq:FormalApprox} into the RHS of 
\eqref{eq:Rrates} for the case of non-overlapping lines in complete 
redistribution.  No preconditioning of the equations is necessary 
due to a serendipitous cancellation of the denominator in the 
source function.  This does not hold, however, for more general 
cases \citep[c.f., the discussion in][]{RybickiHummer91a}.}
\begin{equation}
\label{eq:ALISimp}
\begin{split}
\mathcal{R}_{lj,kd} = \frac{4\pi}{h} 
   \sum_{i=0}^{Q-1} \frac{q_i}{\nu_n}
   \Bigl[
   U_{lj,i}\Bigl(1 - \theta \Lambda_{kdi}^*\Bigr)\\
   + V_{lj,i}\left(u_{kdi}^{\dag} 
   - \theta \Lambda_{kdi}^* S_{kdi}^{\dag}\right)\Bigr].
\end{split}
\end{equation}
To obtain equation~\eqref{eq:ALISimp}, we also neglected 
background and scattering processes in the extinction and emissivity.
We used this form of the ALI system to calculate the results presented 
in \S\ref{sect:Results}.

\section{Results}
\label{sect:Results}

We present results from several calculations using the 
atomic physics and external radiation models described in 
\S\ref{sect:PhysModel}.  We performed a 
total of six calculations: Models I and II test our numerical 
solution against previous work in the literature and explore the approach 
of our solutions toward an independently computed steady-state solution.
Model III represents a canonical solution that exhibits a number of 
important time-dependent effects associated with the transfer of radiation 
through the time-dependent medium.  Finally, Models IV--VI explore 
the consequences of varying certain parameters that characterize the external 
radiation source and atomic physics.  The essential model properties are 
summarized in Table~\ref{tab:Models}, and are described in detail below.

\begin{table}
\caption{Descriptions of the calculations presented in \S\ref{sect:Results} of 
the text.  The canonical model uses the three-level atom described in 
\S\ref{sect:PhysModel}, and the conditions for the external radiation 
set by equation~\eqref{eq:lightcurve}.  The parameter values for the latter 
are given in \S\ref{subsect:Canonical}.  The descriptions of 
the external source and atomic models in the third column 
denote deviations of the calculation from the canonical case.}
\centering
\begin{tabular}{c c c}
Model & External Source & Atomic Model\\
\hline
Model I & Linear increase to $I_{\rm B}$ & Two-level\\
Model II & Linear increase to $I_{\rm B}$ & Canonical\\
Model III & Canonical & Canonical\\
Model IV & Canonical & $A_{21}$ multiplied by ten\\
Model V & $A / I_{\rm o} = 1$ & Canonical\\
Model VI & $A / I_{\rm o} = 1$, $I_{\rm B} / I_{\rm o} = 0.1$ & Canonical\\\hline
\end{tabular}
\label{tab:Models}
\end{table}

\subsection{Comparison to Previous Work and Numerical Tests (Models I and II)}
\label{subsect:Comparison}

As a basic test of our code, we reproduced the results of 
\citet[][]{Kunasz83a}, using the methods of \S\ref{sect:Numerical} 
(Model I).  The earlier work used a 
constant source for the external radiation and 
considered time-dependent radiative 
transfer in a static medium consisting of two-level atoms that 
effectively remained in the ground state.  The line profile for the transition 
was assumed to be a Doppler or Voigt profile; the latter 
used parameter $a = 0.01$.  
For a resonant transition in a two 
level system, a steady-state medium is a good approximation, as each 
excitation event is followed by a spontaneous 
de-excitation to the ground state.  We reproduced this effect 
with our time-dependent methods by restricting our atomic model to 
the $j=0\rightarrow 2$ transition.  In this case, 
the $n_2$ population remains negligible compared to $n_{\rm o}$, 
but the source function for the transition is 
finite.\footnote{That is, $A_{20} n_2 / (B_{02} n_0 - B_{20} n_2)$, 
remains significant, though $n_2 / n_{\rm o} \ll 1$.}
To compare with previous work, we did not use the initial 
conditions described in 
\S\S\ref{subsect:RTESol}--\ref{subsect:RatesSol}, but rather 
set all atoms in the ground state at $t / T_{\rm ref} = 0$.

We performed calculations using this restricted model for 
grid resolutions of $K = 170$, $D = 140$, 
$N = 40$, and $M = 10$.  
The reference depth points were 
spaced with 
$20$ points per decade in $\tau$.  The temporal resolution, 
$K$, was chosen to approximately equal the resolution in 
the spatial grid.
We used the ramp model for white, isotropic external 
radiation at the boundary.
Our time grid was logarithmically spaced over the 
interval $[10^{-4}, 2\times 10^4]$ in units of 
$T_{\rm ref}$.  For the line profile, 
we used a Voigt function with 
parameter $a = 0.01$, and divided the frequency points 
evenly between the linear and logarithmic sections of the grid 
(see \S\ref{subsect:Discretization}).
In the work of \citet[][]{Kunasz83a}, 
the radiation field was separated into unscattered and diffuse 
components, consisting of photons that underwent zero 
and at least one radiative interaction, respectively.  We 
computed the unscattered component, $I_{\nu}^{\rm unsc}$ by fixing the level 
populations at their converged values (which determined the extinction 
coefficient), and then set the source function to zero.  The diffuse field 
was calculated as 
$I_{\nu}^{\rm diff} = I_{\nu} - I_{\nu}^{\rm unsc}$.

\begin{figure}
\begin{center}
\includegraphics[scale=0.40]{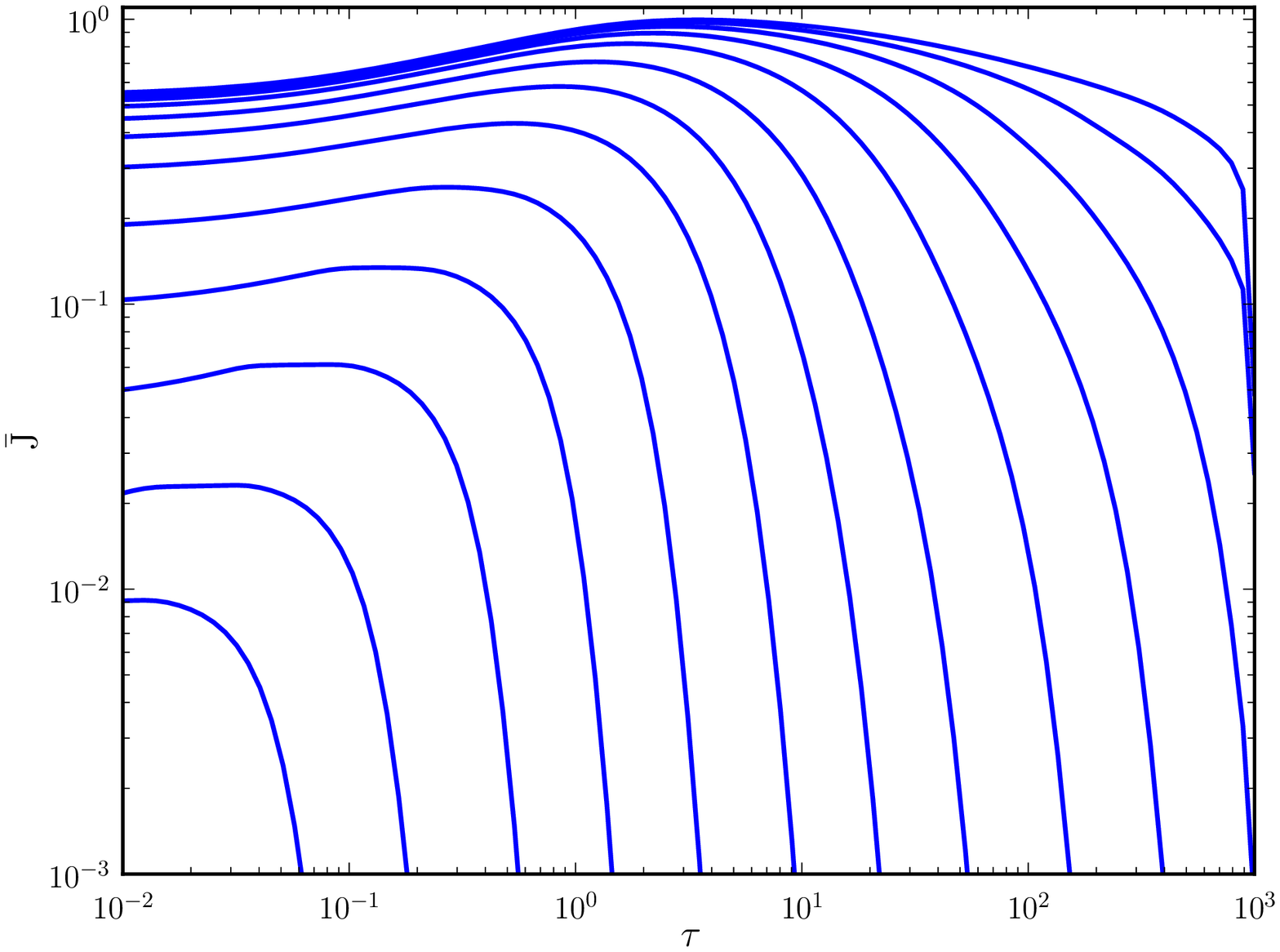}\\
\includegraphics[scale=0.40]{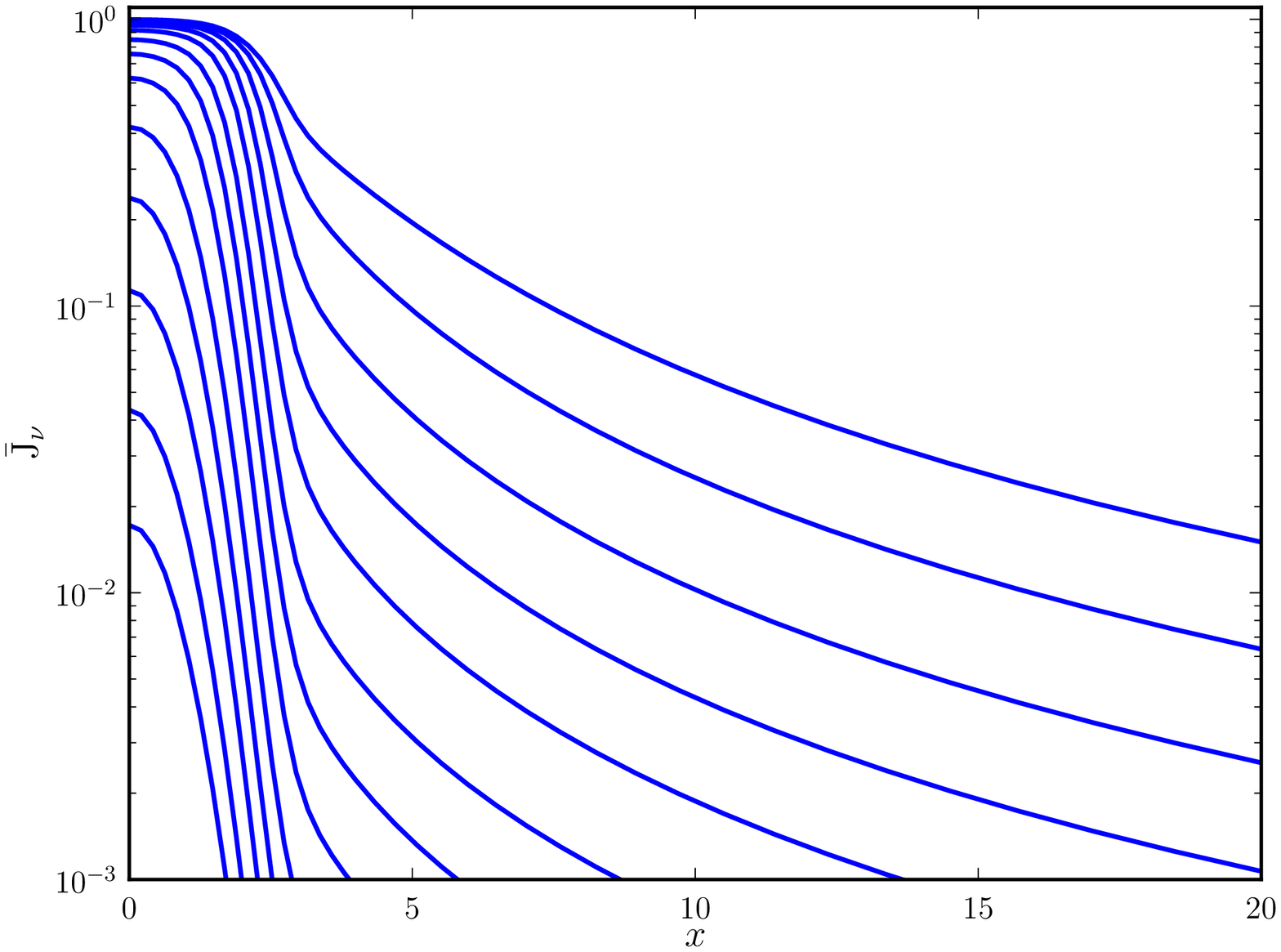}
\end{center}
\caption{Plot of the static, two-level source function, $\bar{\rm J}$
for the $j = 0\rightarrow 2$ 
transition as a function of $\tau$ (top panel); and the angle 
averaged intensity ${\rm J}_{\nu}$ as a function of frequency in 
Doppler widths, $x$ (bottom panel).  In the bottom plot, 
the results are 
shown at $\tau = 0$.  
Calculations were performed for Model~I (see \S\ref{subsect:Comparison}).
For both plots, each curve represents a 
distinct time, increasing monotonically from bottom to top.  
The curves are all normalized to the maximum value of an independently 
calculated steady-state solution 
(see \S\S\ref{subsect:RTESol}--\ref{subsect:RatesSol}).
The figure may be compared with 
the righthand top and bottom panels of Fig.~1 
in \citet[][]{Kunasz83a}.}
\label{fig:compare}
\end{figure}

Figure~\ref{fig:compare} shows the results of our calculations, which 
can be 
compared with the righthand top and bottom panels of Fig.~1 in 
\citet[][]{Kunasz83a}.  The top panel shows the two-level source 
function for the diffuse component, 
$\bar{\rm J} = (4\pi)^{-1} \int_{4\pi} d\Omega \int d\nu \varphi_{\nu} 
 I_{\nu}^{\rm diff}$, as a function of reference depth, $\tau$, at 
several times.  The bottom panel shows the angle-averaged intensity, 
${\rm J}_{\nu} \equiv (4\pi)^{-1} \int_{4\pi} d\Omega I_{\nu}^{\rm diff}$, 
at $\tau = 0$, as a function of $x$, for several times.  
The curves are all normalized to the maximum value of an independently 
calculated steady-state solution 
(see \S\S\ref{subsect:RTESol}--\ref{subsect:RatesSol}).
The different curves 
correspond to monotonically increasing values of the time from bottom to top 
in each panel.  
These times represent the closest values on our grid to those listed in 
Table~1 of \citet[][]{Kunasz83a}.  To the extent we were able to 
compare, our results are in excellent agreement with the earlier work, 
reproducing the main features and trends, and showing only slight 
quantitative disagreement in one of the plots.\footnote{In the righthand 
bottom panel of Figure~1 in \citet[][]{Kunasz83a}, we obtain quantitative 
agreement if the labelling of curves `b' -- `k' is shifted down by one 
letter to `a' -- `j'.}
Though we show a comparison to only two figures for brevity, we are able to 
reproduce all of the essential results of \citet[][]{Kunasz83a}.

The main features of Figure~\ref{fig:compare} are propagation of the 
external radiation through the medium with increasing time in the top panel, 
and the approach of the curves toward an independently calculated steady 
state solution 
in both panels as $t / T_{\rm ref}$ increases beyond the 
light crossing time $T_{\rm LC}$.

Using the same model for the external source, we performed a similar 
calculation using the three-level atomic model described in 
\S\ref{sect:PhysModel}, and a Voigt line profile (Model II).  In this case, 
the medium is no 
longer approximately static, leading to a non-negligible population in 
the first excited state, $j = 1$ (see \S\ref{subsect:Canonical}). 
This leads to a difference in the source functions for the 
$j = 0\rightarrow 2$ and $j = 1\rightarrow 2$ transitions compared to 
the two-level case, altering the diffuse radiation field in the medium.  

\begin{figure}
\begin{center}
\includegraphics[scale=0.40]{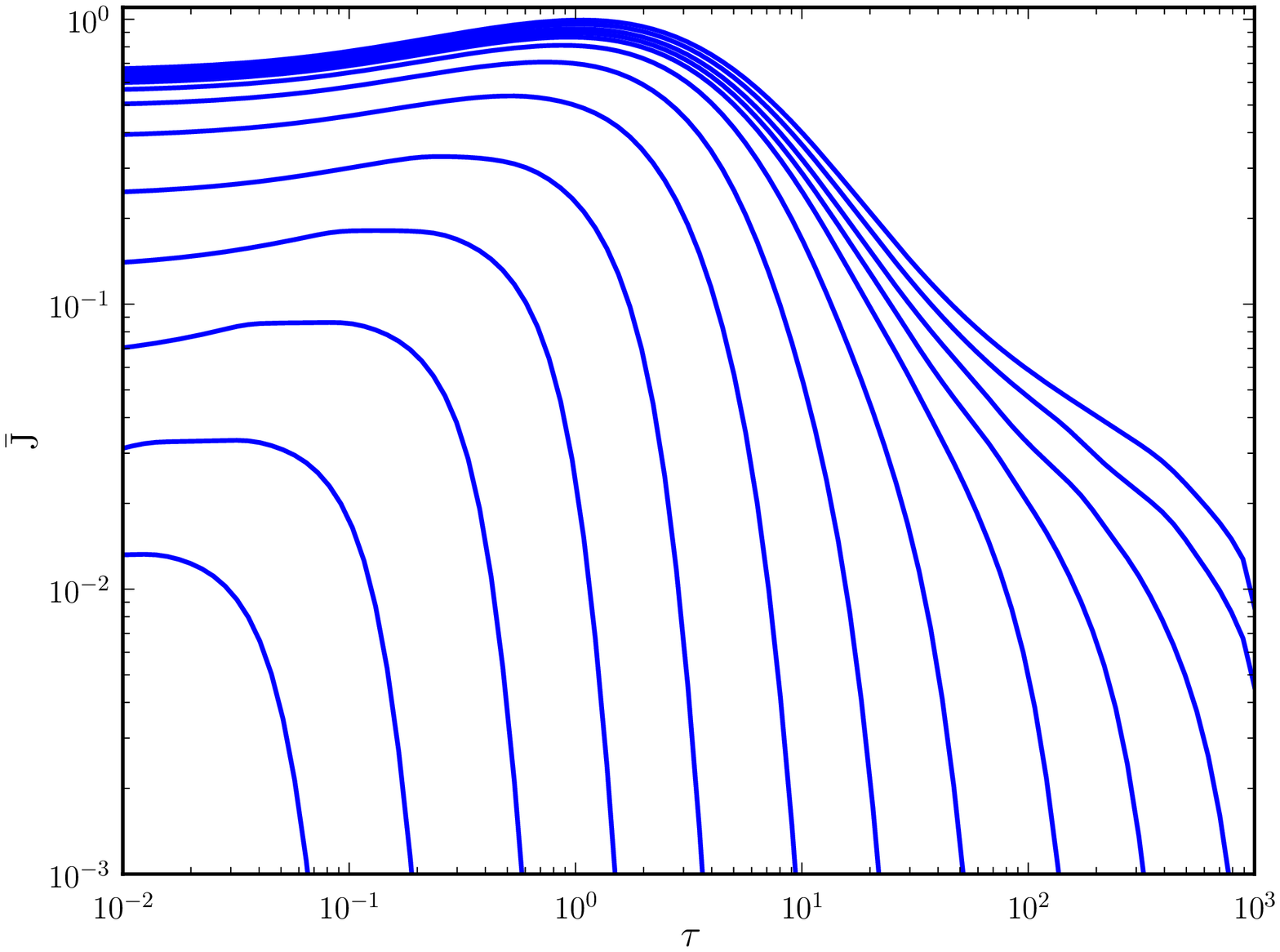}\\
\includegraphics[scale=0.40]{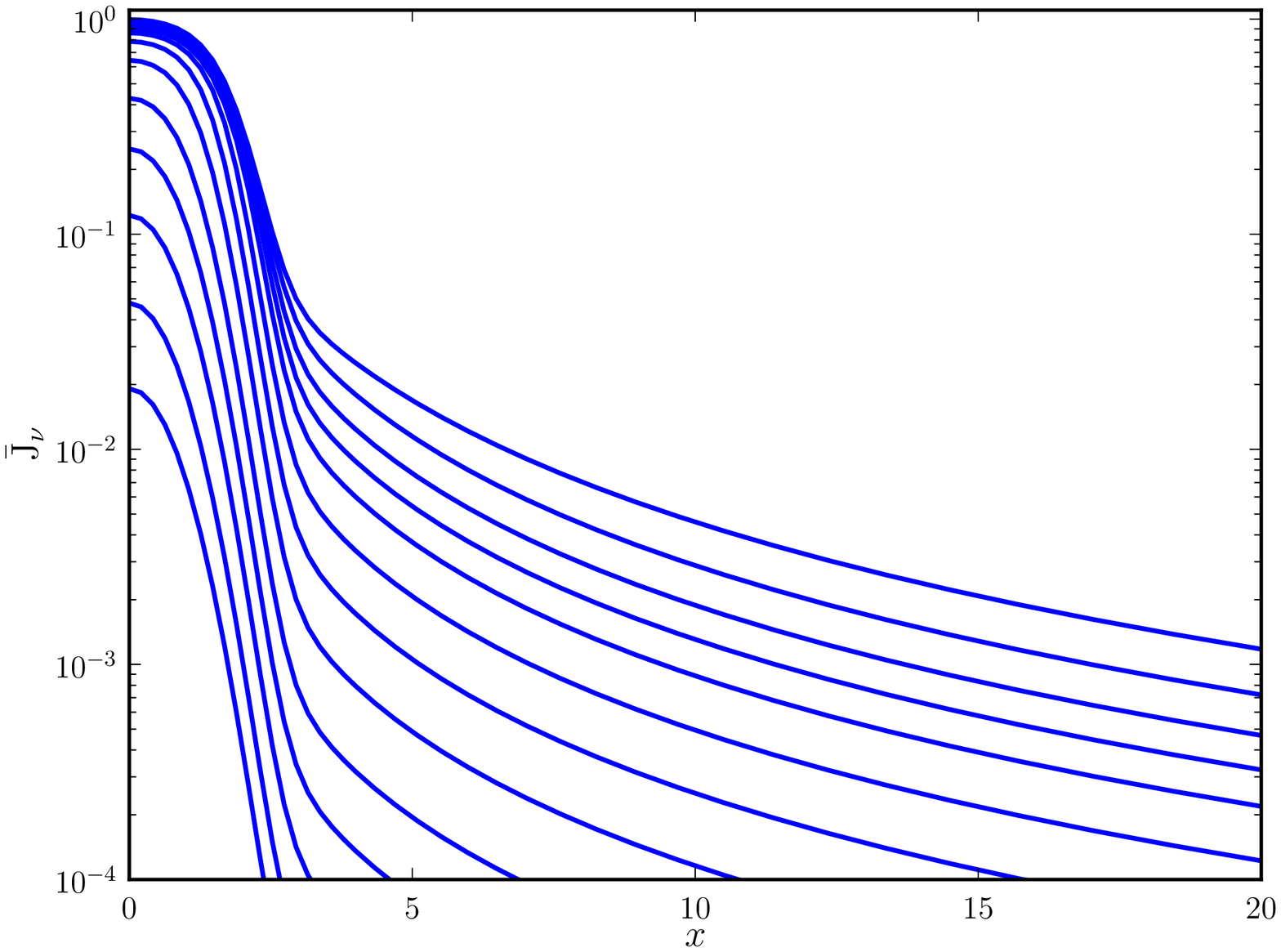}
\end{center}
\caption{Same as Figure~\ref{fig:compare}, except for the three 
level atomic model (Model~II of \S\ref{subsect:Comparison}).  
The quantities shown were
computed using the radiation 
field within $x\le 20$ of line centre for the 
$j = 0\rightarrow 2$ transition.}
\label{fig:tlt_02}
\end{figure}

Figure~\ref{fig:tlt_02} is identical to Figure~\ref{fig:compare}, except that 
it shows results for the three-level model, using the radiation field 
within $x\le 20$ of line centre for the $j = 0\rightarrow 2$ transition.  
With this 
atomic model, the radiation field exhibits a stronger depression at large 
reference depths, as well as a slightly softer line profile at $\tau = 0$.  
This is caused by the fact that, in contrast to the static two-level model, 
not every excitation to $j = 2$ results in a decay to $j = 0$; 
a fraction of de-excitations are to the $j = 1$ level, determined by the 
values of the populations and the Einstein coefficients for each transition.
An important feature of Figures~\ref{fig:compare} and \ref{fig:tlt_02} is 
that the magnitude of the angle-averaged diffuse field becomes a significant 
fraction of the external $J_{\nu}$ only after an interval $\gtrsim T_{\rm ref}$.  
This is expected, as 
$T_{\rm ref}$ represents 
the characteristic time interval between successive radiative interactions.  
This point is important for judging whether a given astrophysical system will 
exhibit time coupled effects in radiative transfer (see 
\S\ref{sect:Discussion}).

\begin{figure}
\begin{center}
\includegraphics[scale=0.40]{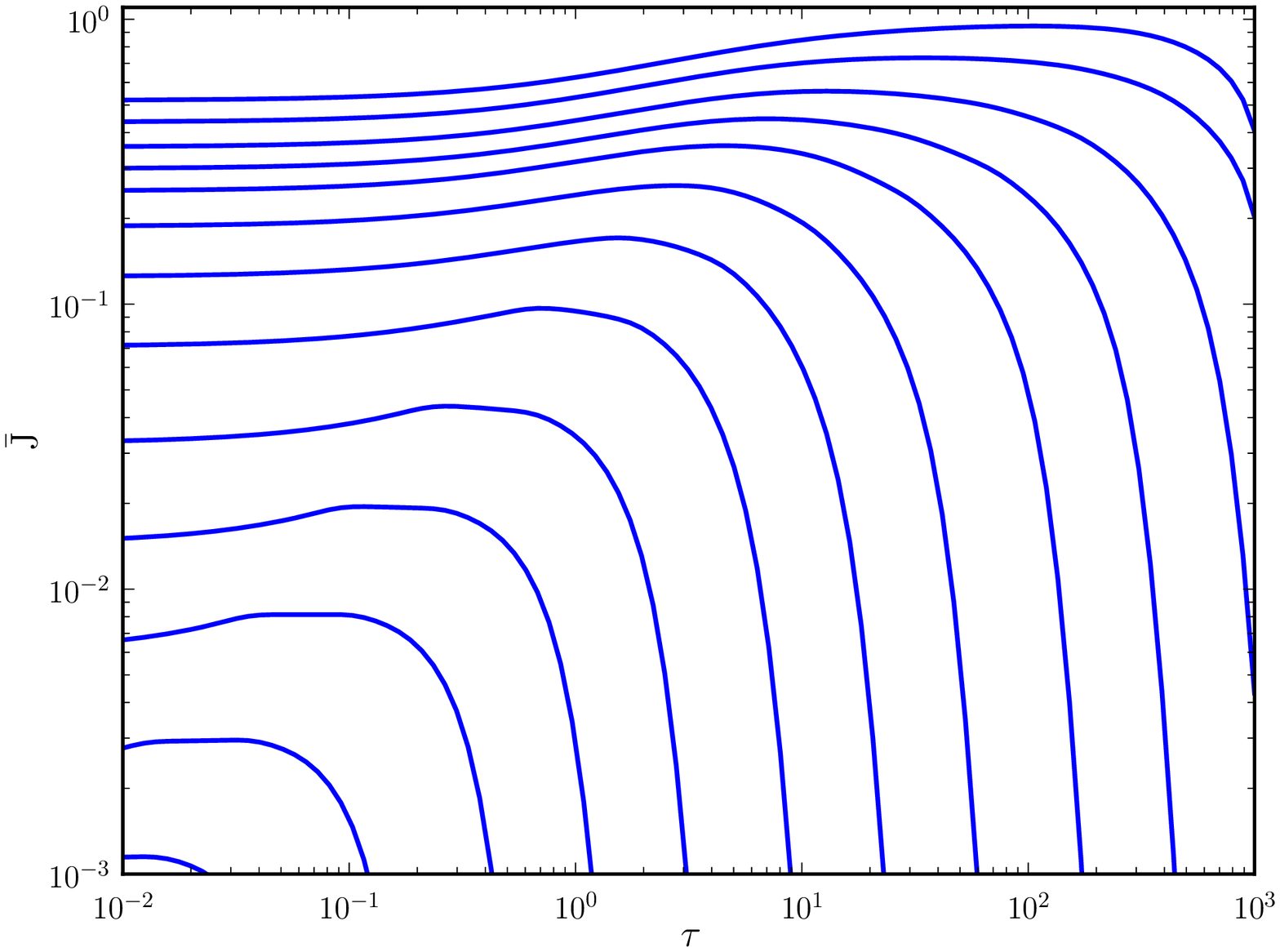}\\
\includegraphics[scale=0.40]{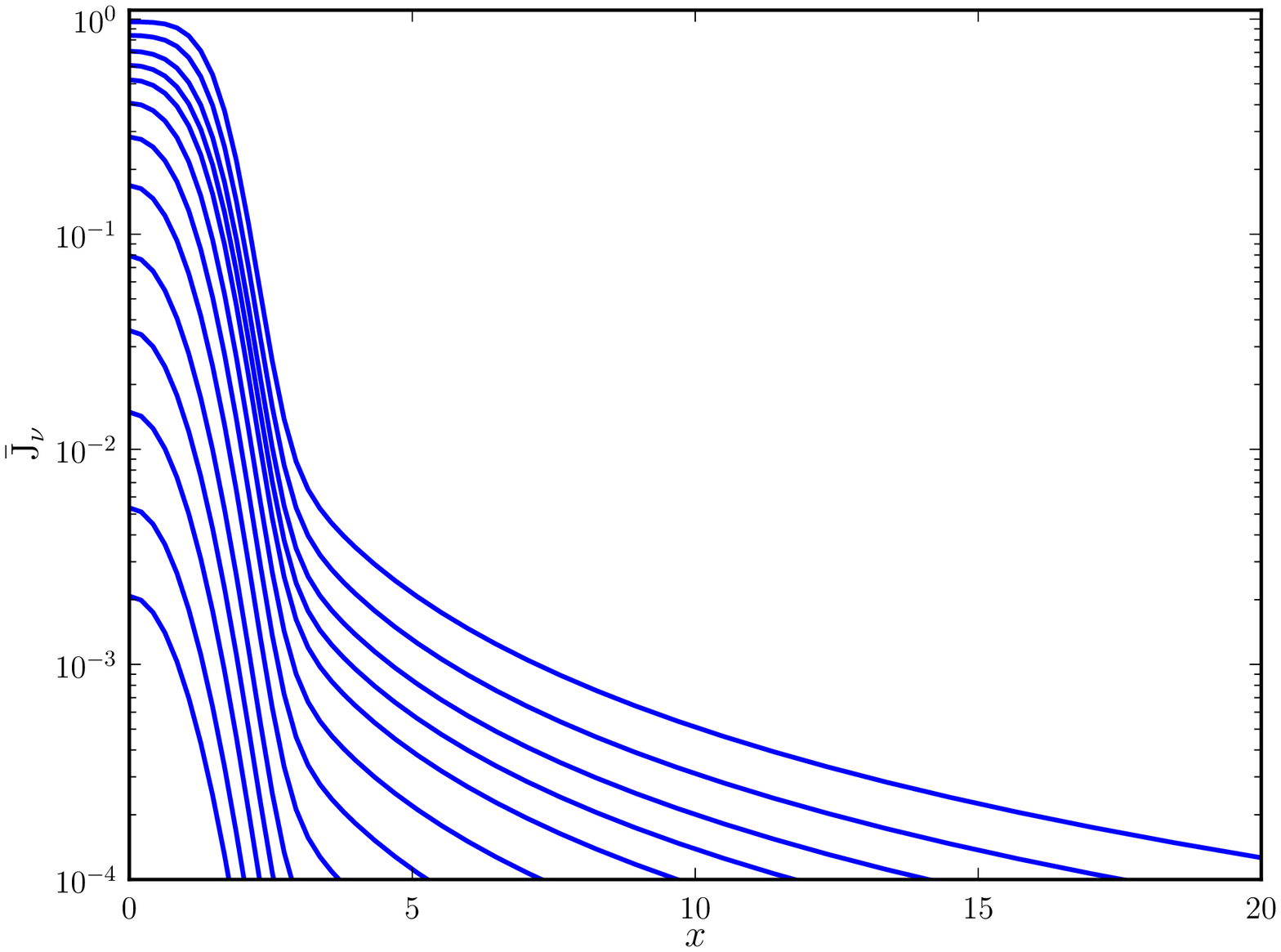}
\end{center}
\caption{Same as Figure~\ref{fig:tlt_02}, except results are shown 
for the radiation field within $x\le 20$ of 
line centre for the $j = 1\rightarrow 2$ transition.}
\label{fig:tlt_12}
\end{figure}

Figure~\ref{fig:tlt_12} is identical to Figure~\ref{fig:tlt_02}, except that 
it shows results at frequencies $x\le 20$, measured from line centre 
of the $j = 1\rightarrow 2$ transition.  The results are similar to those 
described above, except that the radiation experiences relatively smaller 
extinction at large reference depth.  This is due to the fact that the 
effective optical depth for the excited state transition is much smaller than 
that for the ground state resonant transition (see \S\ref{subsect:Canonical}).
As in Model I, the results for Model II converge to an independently 
calculated steady-state as $t$ becomes larger than $T_{\rm LC}$.  

\S IIId and \S IV of \citet[][]{Kunasz83a} present a detailed analysis of the 
method of lines solution to the radiative transfer equation in a two-level, 
static medium, analysing both 
its stability and effectiveness for various test problems.  We have come to 
similar conclusions regarding our solution for the three-level, time-dependent 
medium.  Briefly, we note that for $\bar{\theta} \lesssim 0.6$, unphysical 
oscillations can be introduced into the solution (though this can be mitigated 
somewhat by reducing the time and space grid spacings).  Also, as the time and 
space grid resolution is increased, the results approach a final, converged 
solution with respect to the grid spacing.  As in previous works, we find that 
distinct spacing in the time and space grids leads to an approximate 
representation of the propagation of the leading front of the external 
radiation.  
As long as the time interval for integration is not too large 
compared to $\tau_{\rm max} T_{\rm ref}$, we can use a linear time grid of 
sufficiently small spacing to provide an adequate sampling of the variation of 
the radiation field at all reference depths.

\begin{figure}
\begin{center}
\includegraphics[scale=0.40]{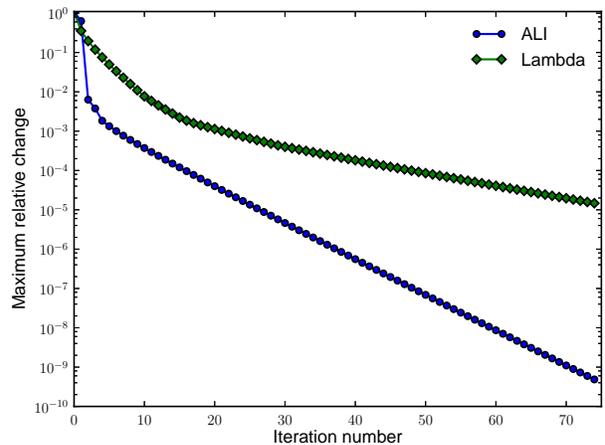}
\end{center}
\caption{Maximum relative change in the level populations as a function of 
iteration number for the ALI and Lambda iteration methods 
applied to Model II of \S\ref{subsect:Comparison}.}
\label{fig:iteration}
\end{figure}

Apart from the solution to the RTE itself, the central concern of our method  
is the convergence of the atomic level populations.  Figure~\ref{fig:iteration} 
shows the maximum relative change in the populations of levels $n_0$, 
$n_1$, and $n_2$, for all time and space grid points, at each iteration 
during the solution.  The curves marked by diamond and circle shaped symbols 
show the relative population change when Lambda and ALI iteration is used, 
respectively.  As expected, the ALI method has a steeper slope and converges 
more quickly for a given number of iterations than the Lambda method.  
The results shown in Figure~\ref{fig:iteration} are fairly typical for the 
models presented in this paper; in fact, the convergence properties are 
often better for Models III--VI than for Model II.  For practical 
implementation of our methods with more complicated atomic models, it will  
likely prove useful to supplement the ALI iteration with additional 
acceleration methods, for example, Ng's method or the generalized minimum 
residual method
\citep[for a detailed discussion of acceleration techniques, 
see, e.g.,][]{Auer91a}.

As we increased the spatial grid resolution, the number of required 
iterations to reach a given relative accuracy also increased; this is a well 
known phenomenon \citep[see][]{Olsonetal86a}.  However, this was typically 
not true of the time grid resolution.
As discussed in \S\ref{subsect:ALI}, the initial guess for the 
solution at each time step is the converged solution for the previous time 
step.  The 
finer the time grid resolution, the more similar the solution is in 
consecutive steps.  Thus, increasing the time grid resolution tends to reduce 
the number of iterations per step.  Since a reduction in spatial grid 
spacing is typically accompanied by a reduction in time grid spacing, 
the two effects will tend to counteract 
one another, and the increase in spatial grid resolution will not adversely 
affect the required number of iterations as much as expected.  
This is an unexpected benefit 
from applying the ALI method to the time-dependent problem.

\subsection{Canonical Transient Pulse (Model III)}
\label{subsect:Canonical}

A class of external sources of immediate interest for this paper are 
those exhibiting transient behaviour, in which the amplitude of the 
transient pulse, $A$, is much greater than the background intensity 
$I_{\rm B}$.
We constructed a canonical model for 
transient sources, using equation~\eqref{eq:lightcurve}, with parameter 
values $I_{\rm B} / I_{\rm o} = 10^{-5} $, $A / I_{\rm o} = 0.1 $, 
$t_0 / T_{\rm ref} = 5\times 10^2 $, and $\sigma = 0.5$ (Model III).  We 
assumed
that the external source was collimated, according to the prescription 
of \S\ref{subsect:Discretization}.
In our calculations, we used grid resolutions of 
$K = 280$, $D = 280$, $N = 20$, and $M = 40$.  The time grid was 
spaced linearly over the interval $[0, 4 T_{\rm LC}]$.

For our atomic model, we used the three-level system described in 
\S\ref{sect:PhysModel}, with a common, Doppler line profile for 
all three transitions.  The initial condition for the radiation 
field and atomic level populations was the steady-state solution 
at $t / T_{\rm ref} = 0$
(see \S\S\ref{subsect:RTESol}--\ref{subsect:RatesSol}). 

\begin{figure}
\begin{center}
\includegraphics[scale=0.40]{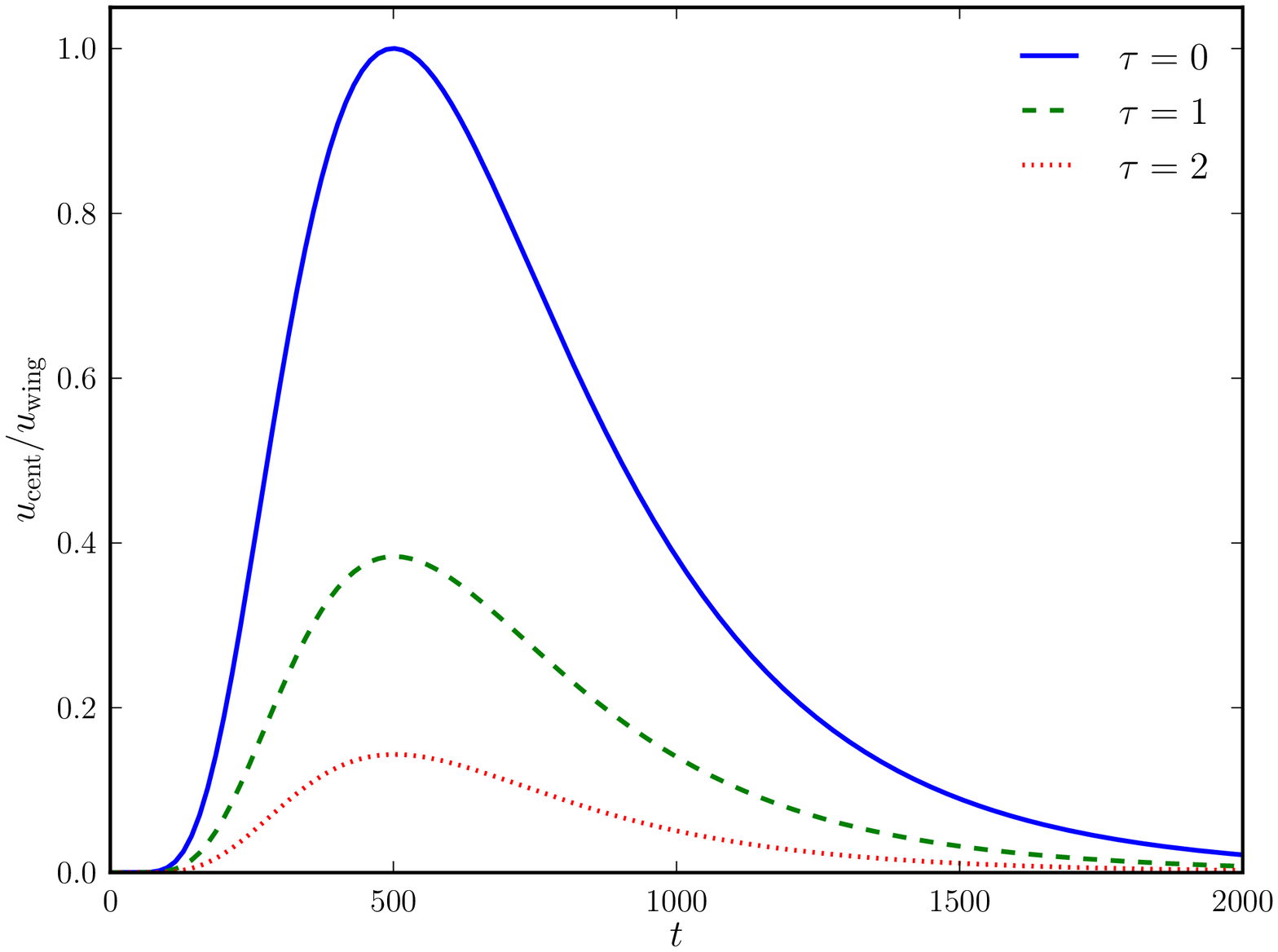}\\
\includegraphics[scale=0.40]{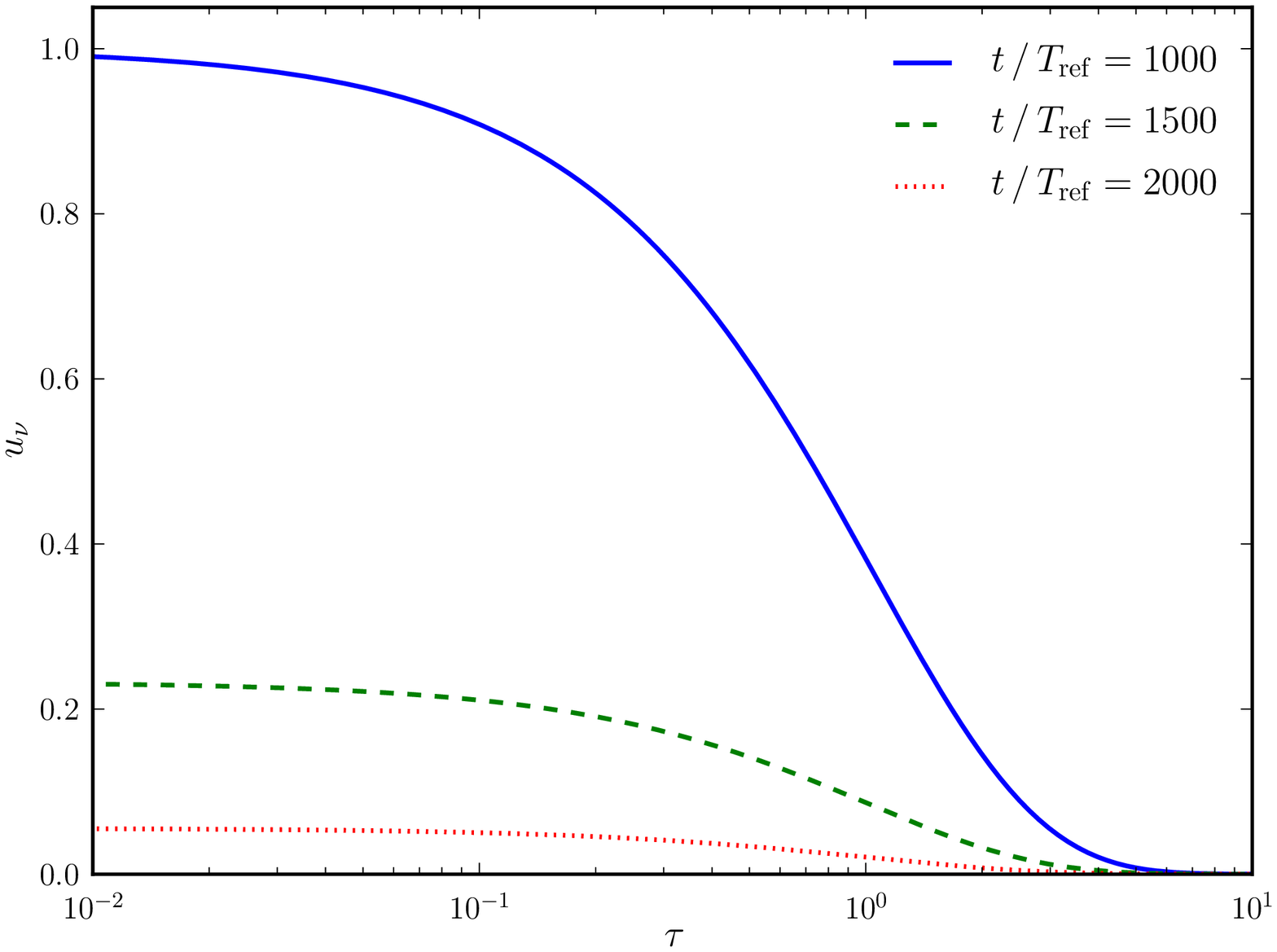}
\end{center}
\caption{Radiation field for the  
$j = 0\rightarrow 2$ transition with $\mu = 1$.  The top panel 
shows the line decrement $u_{\rm cent} / u_{\rm wing}$ (see the 
definition in \S\ref{subsect:Canonical}) 
for several values of the 
reference depth.  The bottom panel shows the radiation 
variable $u_{\nu}$ as a function of reference depth at several times.  
In the bottom panel, the variable $u_{\nu}$ is normalized to its maximum 
value in the $t / T_{\rm ref} = 1000$ curve.
The time is measured in units of $T_{\rm ref}$.
All calculations have been 
performed for the canonical Model III of \S\ref{subsect:Canonical}.}
\label{fig:u02_MIII}
\end{figure}

We define the decrement in the radiation field associated 
with a given transition as the 
value of $u_{\nu}$ at line centre ($x = 0$) divided by the value in 
the line wing ($x = 4$). The decrement is 
denoted $u_{\rm cent} / u_{\rm wing}$.
Figure~\ref{fig:u02_MIII} shows the results of our calculation for the 
the $j = 0\rightarrow 2$ transition, with 
direction $\mu = 1$.
The top panel shows the line decrement as a function of 
$t$ at several reference depths, $\tau = 0, 1, 2$.
The time $t$ is measured 
in units of $T_{\rm ref}$.
Larger depths are not plotted, as the radiation field is negligible 
due to large extinction at $\tau \gg 1$.  The peak intensity occurs at 
$t / T_{\rm ref}\approx 500$, when the external radiation has its maximum 
value.  At $\tau = 0$, there is little radiation re-emitted in the 
$\theta > 0$ direction, implying 
that $u_{\nu}\approx I_{\nu}^- / 2 \approx I_{\nu}^{\rm ext} / 2$.
The bottom panel of Figure~\ref{fig:u02_MIII} shows $u_{\nu}$ 
as a function of $\tau$ for several times:
$t / T_{\rm ref} = 1000, 1500, 2000$.  These times correspond to $T_{\rm LC}, 
1.5 T_{\rm LC}$, and $2 T_{\rm LC}$, respectively.  
The curves in the bottom panel are normalized to the maximum value of 
$u_{\nu}$ at $t / T_{\rm ref} = 1000$.

\begin{figure}
\begin{center}
\includegraphics[scale=0.40]{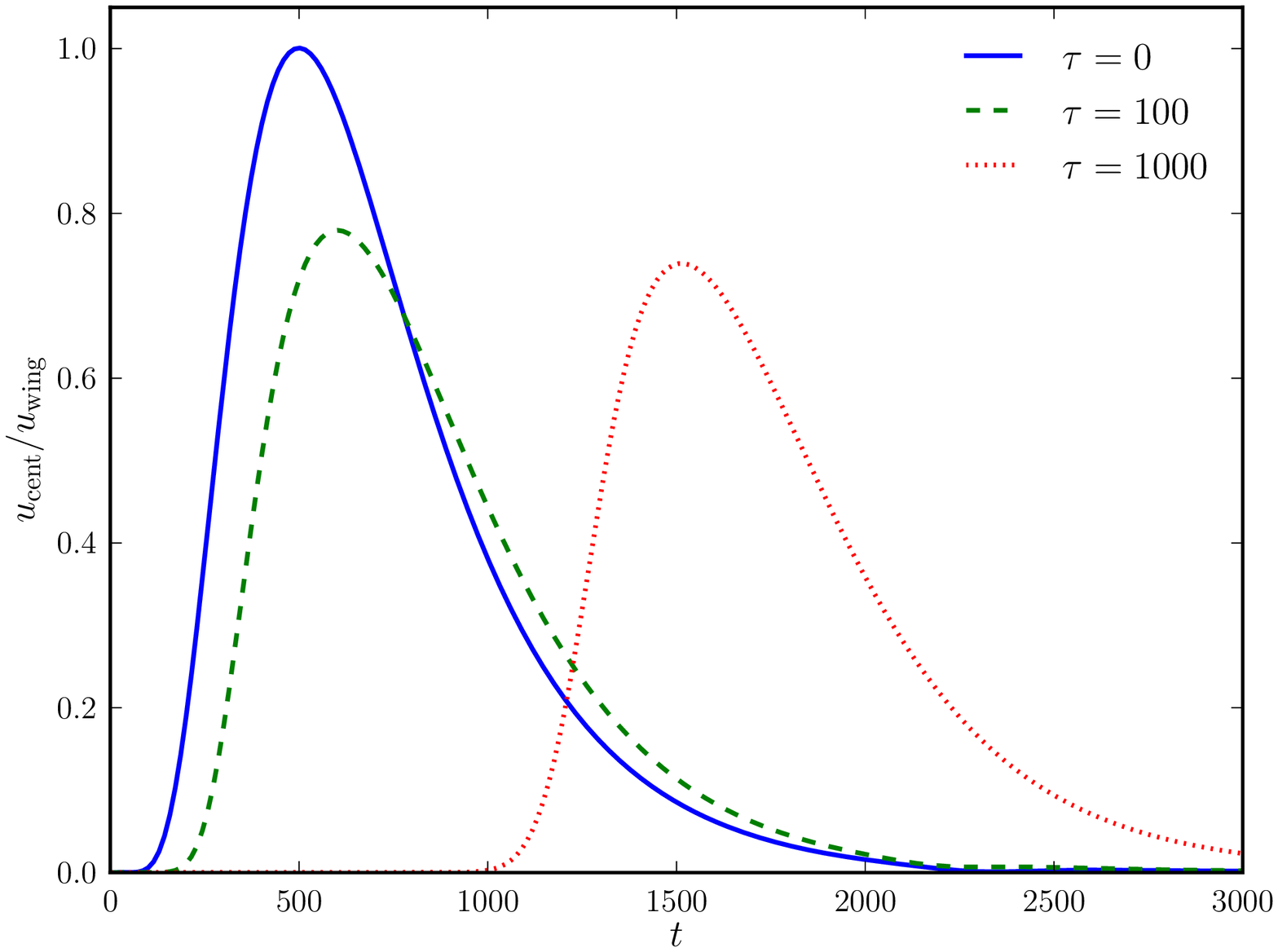}\\
\includegraphics[scale=0.40]{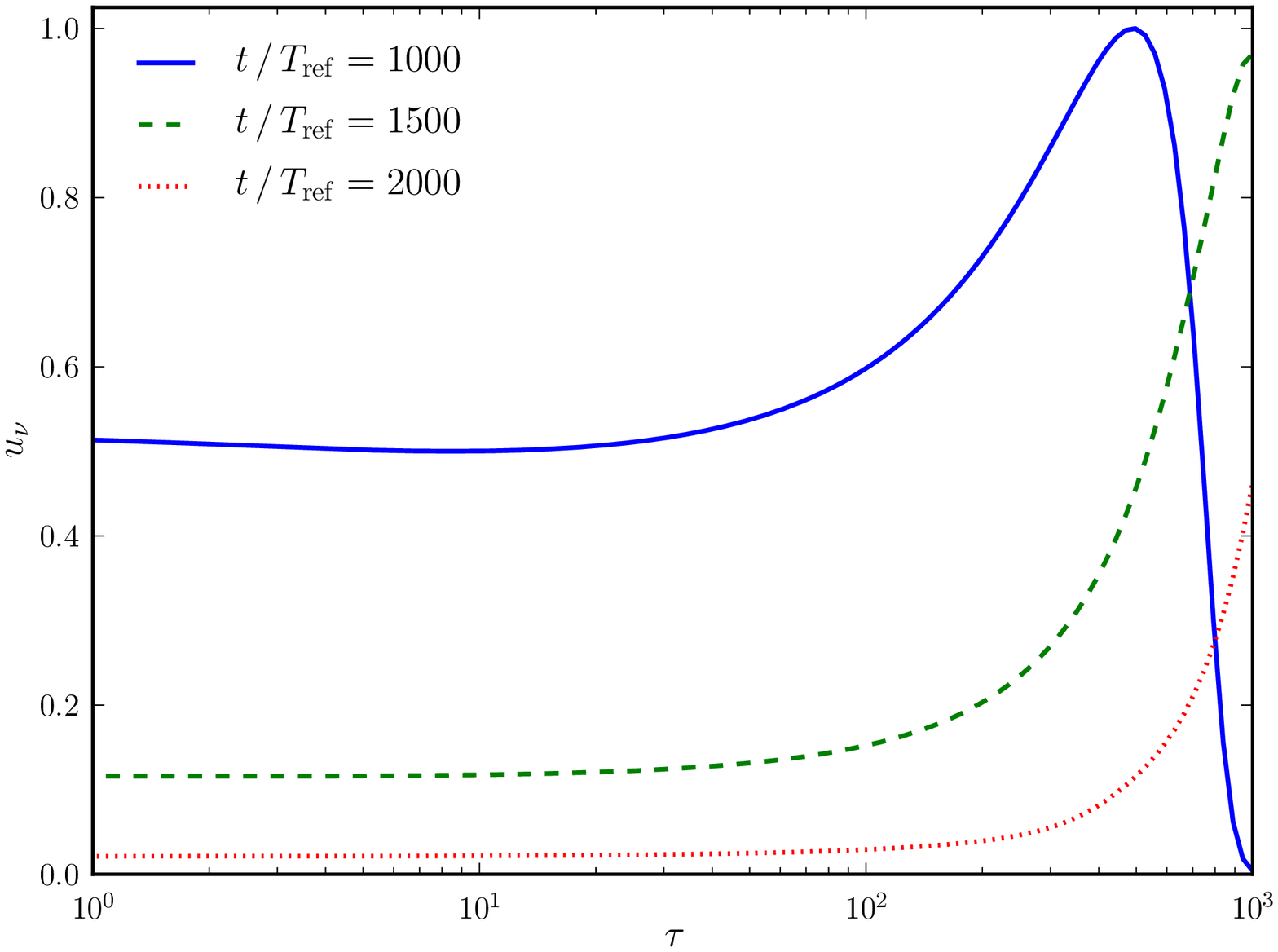}
\end{center}
\caption{Same as Figure~\ref{fig:u02_MIII}, except results 
are shown for the 
$j = 1\rightarrow 2$ transition at reference optical depths 
$\tau = 0, 100, 1000$.}
\label{fig:u12_MIII}
\end{figure}

Figure~\ref{fig:u12_MIII} is similar to Figure~\ref{fig:u02_MIII}, except 
that it shows results for the $j = 1\rightarrow 2$ transition.
The radiation field is non-zero at larger
reference depths than in the transition $j = 0\rightarrow 2$ because the 
effective 
optical depth for the $j = 1\rightarrow 2$ transition is 
smaller.
This is due to the reduced population of the excited state relative to 
the ground state, as well as the smaller Einstein absorption coefficient for 
the $j = 1\rightarrow 2$ transition compared to the resonant transition. 
It should be noted that most of the extinction in this transition 
occurs for $\tau < 100$.  The radiation field at frequencies near the 
resonant transition, which is necessary for the excitation event  
$j\rightarrow 2\rightarrow 1$, is largely damped at $\tau = 100$. 
Thus, 
there are few excitation events to $j = 1$ at $\tau > 100$, 
and therefore little difference in the extinction from the 
$j = 1\rightarrow 2$ transition at $\tau = 100$ versus 
$\tau = 1000$.

The smaller extinction is also evident in the bottom panel of 
Figure~\ref{fig:u12_MIII}, in which the crossing of the spatial 
radiation profile through the $\tau = \tau_{\rm max}$ boundary is evident 
as time increases.
As expected, the maximum of the radiation pulse is shifted 
to later times at larger reference depths.  This is due to the fact that 
the light crossing time is of order the variability timescale for the 
radiation source.  This feature is absent from Figure~\ref{fig:u02_MIII} 
because the reference depths plotted are small compared to 
$T_{\rm LC} / T_{\rm ref}$.

The results for the radiation field at frequencies around line centre of the 
$j = 0\rightarrow 1$ transition are similar to those shown in 
Figure~\ref{fig:u12_MIII}, except for the absence of significant extinction, 
due to the small value of the Einstein absorption coefficient for this 
transition.

\begin{figure}
\begin{center}
\includegraphics[scale=0.40]{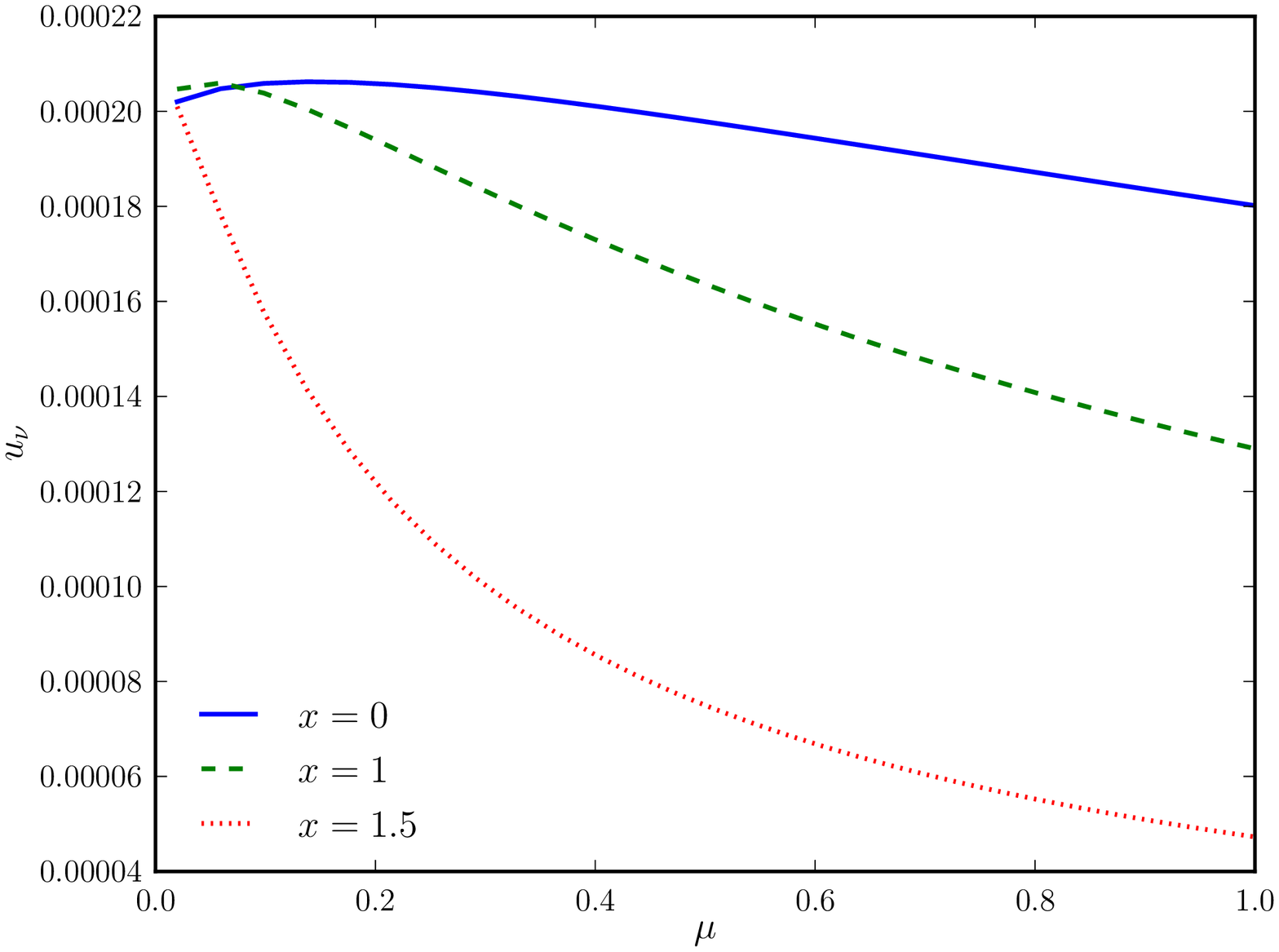}\\
\includegraphics[scale=0.40]{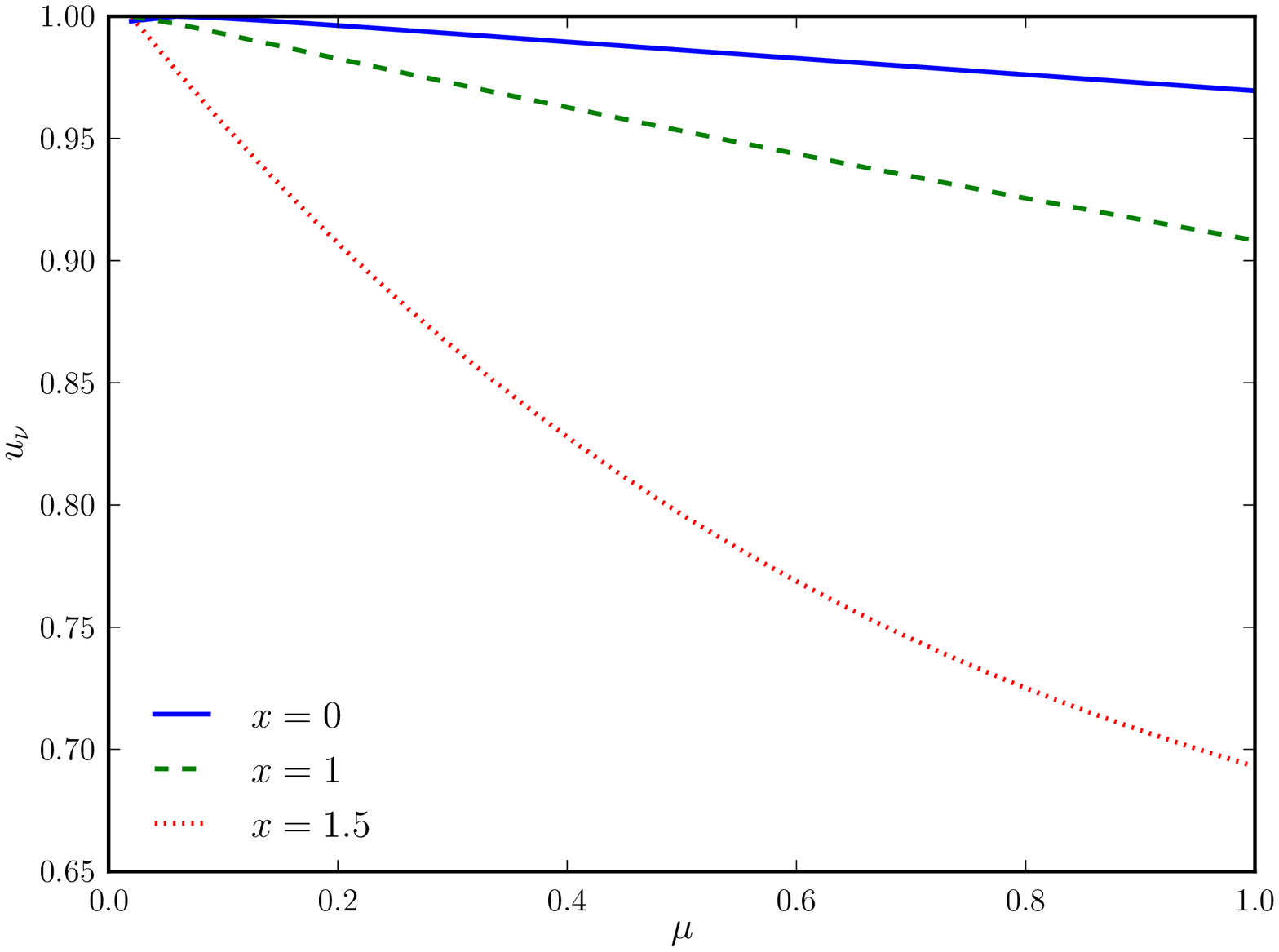}
\end{center}
\caption{Radiation variable $u_{\nu}$ as a function of 
$\mu$ for several values of frequency in units of 
Doppler width, $x$.  The top and bottom panels show results at 
$\tau = 0$ and $\tau = \tau_{\rm max}$ for times 
$t / T_{\rm ref} = 500$ and $t / T_{\rm ref} = 500 + \tau_{\rm max}$, 
respectively.  All calculations 
were performed for the canonical Model III of \S\ref{subsect:Canonical}.}
\label{fig:u02th_MIII}
\end{figure}

Figure~\ref{fig:u02th_MIII} shows the angular profile of the radiation at 
several frequencies around line centre of the $j = 0\rightarrow 2$ 
transition.  Results are plotted at $x = 0, 1,$ and $1.5$.  
The top panel shows the values of the $u_{\nu}$ radiation field variable 
at $\tau = 0$ for $t / T_{\rm ref} = 500\,T_{\rm ref}$, when the external 
source peaks at the inner boundary.  The bottom panel shows the values of 
$u_{\nu}$ at $\tau = \tau_{\rm max}$ for $t / T_{\rm ref} = 1500\,T_{\rm ref}$, 
when the peak of the external radiation passes through the outer boundary.  
We show the results for the diffuse field only; therefore, 
values at $\mu = 1$ are excluded from the plots.
All curves are normalized to the maximum value of $u$ at each 
frequency $x$.  The plots demonstrate that the radiation field remains 
highly collimated at the $\tau = 0$ boundary  
but is largely isotropic at $\tau_{\rm max}$ except at $x > 1$.
This result makes sense: because the optical depth 
in the $j = 0\rightarrow 2$ transition is large, photons that reach 
$\tau = \tau_{\rm max}$ have 
experienced many atomic interactions and 
have been redistributed isotropically in angle.  By contrast, 
the angular distribution of photons at the $\tau = 0$ boundary is 
largely set by the external radiation;
only 
backscattered photons from nearby layers of the slab are redistributed 
in angle at $\tau = 0$.  The medium therefore transitions from highly 
collimated to largely isotropic with intermediate results 
at $\tau \gtrsim 1$.

\begin{figure}
\begin{center}
\includegraphics[scale=0.40]{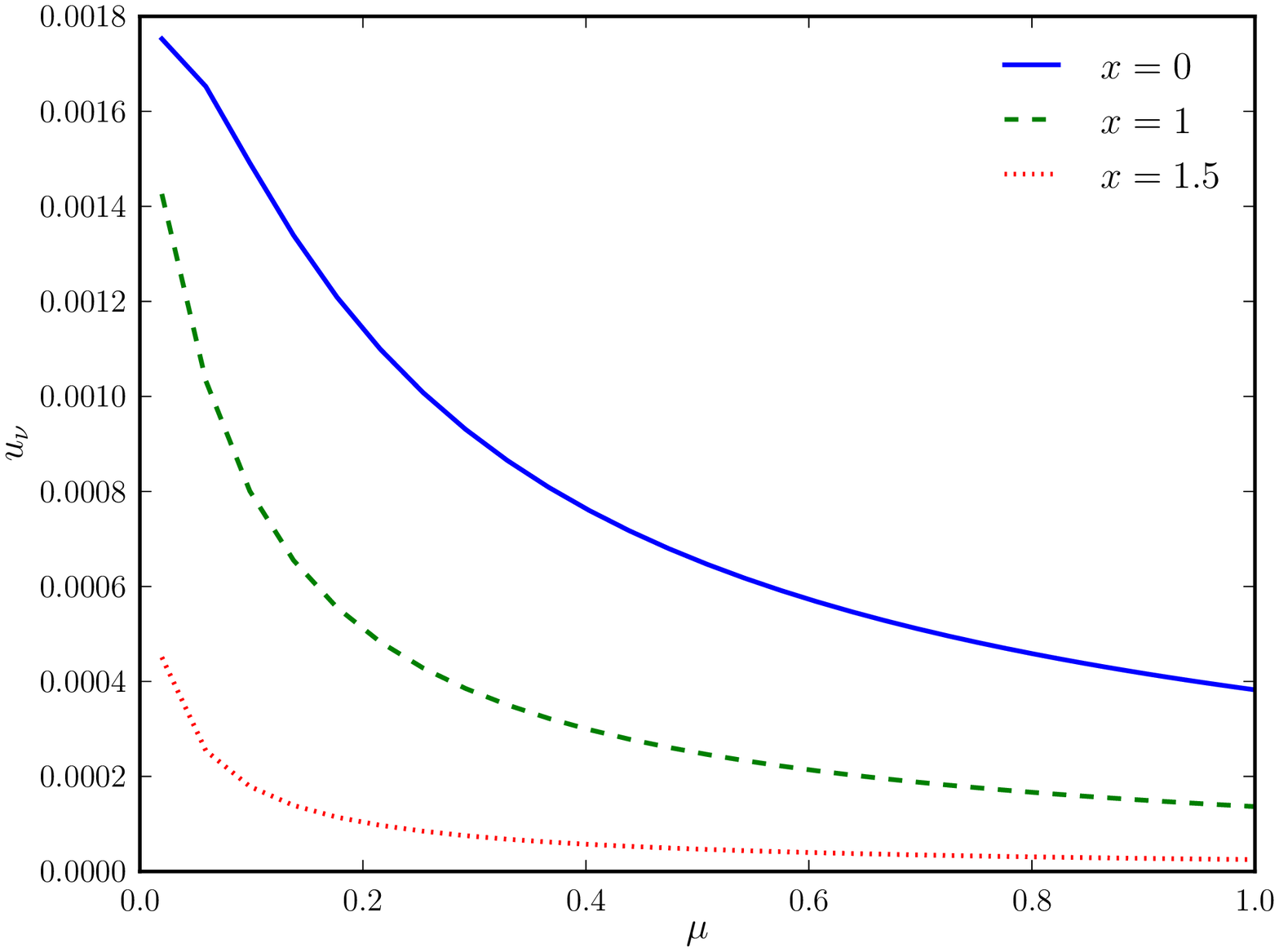}\\
\includegraphics[scale=0.40]{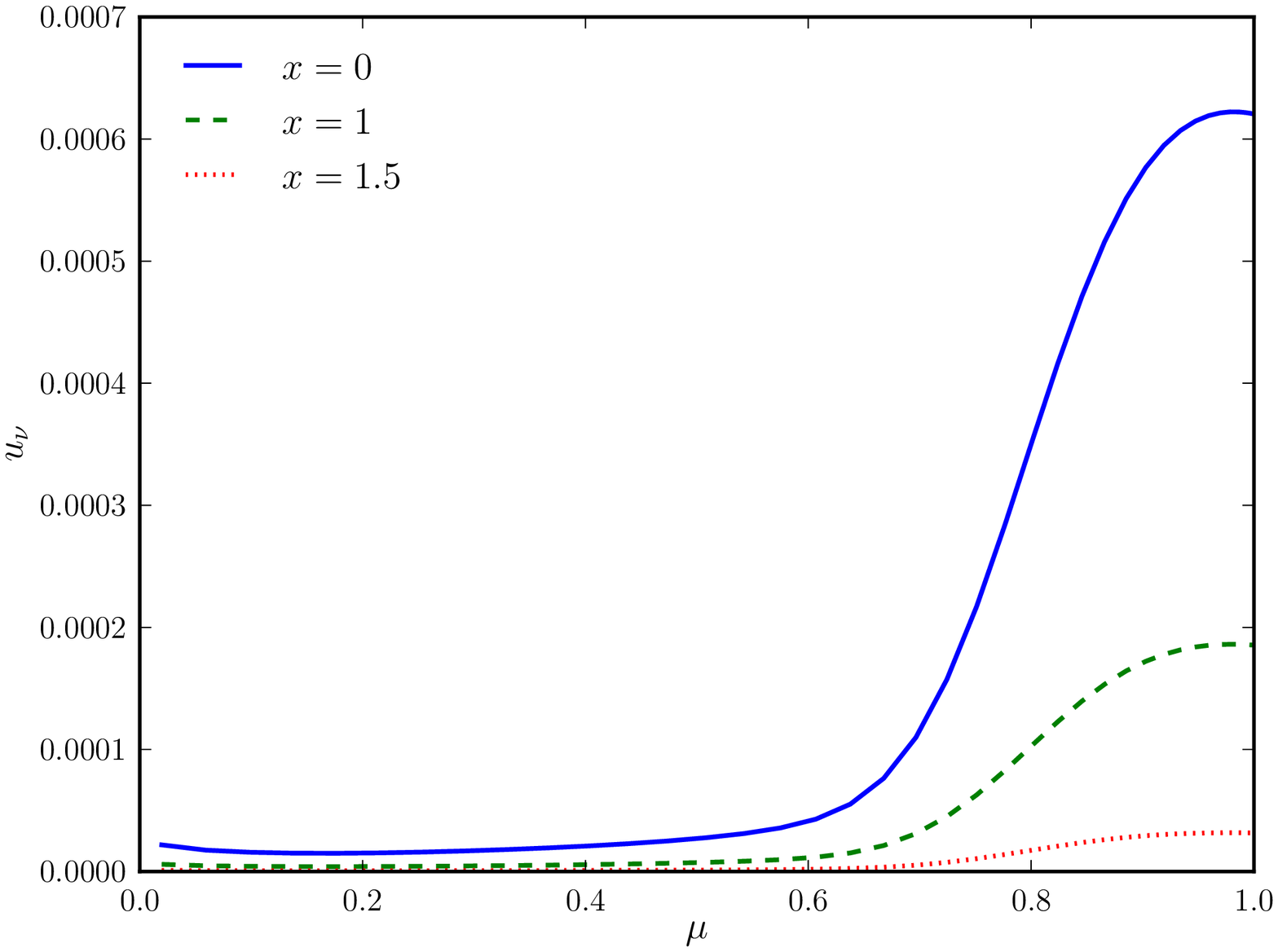}
\end{center}
\caption{Same as Figure~\ref{fig:u02th_MIII}, except for frequencies 
around the $j = 1\rightarrow 2$ transition.}
\label{fig:u12th_MIII}
\end{figure}

Figure~\ref{fig:u12th_MIII} is the same as Figure~\ref{fig:u02th_MIII} 
except that it shows results for frequencies around the $j = 1\rightarrow 2$ 
transition.  
In both cases, the radiation with $\theta > 0$ is a small fraction  
of $u_{\nu}$ at $\tau = 0$.  However, for the $j = 1\rightarrow 2$ transition, 
the radiation field remains highly anisotropic even at $\tau = \tau_{\rm max}$. 
This is due to the fact that this transition has a small effective optical depth 
(see below).  Therefore, a small fraction of 
photons reaching $\tau = \tau_{\rm max}$ have been absorbed and redistributed 
into angles $\mu\ne 1$.  
The radiation in the $j = 0\rightarrow 1$ transition exhibits 
a nearly identical angular profile to the $j = 1\rightarrow 2$ 
transition.

\begin{figure}
\begin{center}
\includegraphics[scale=0.40]{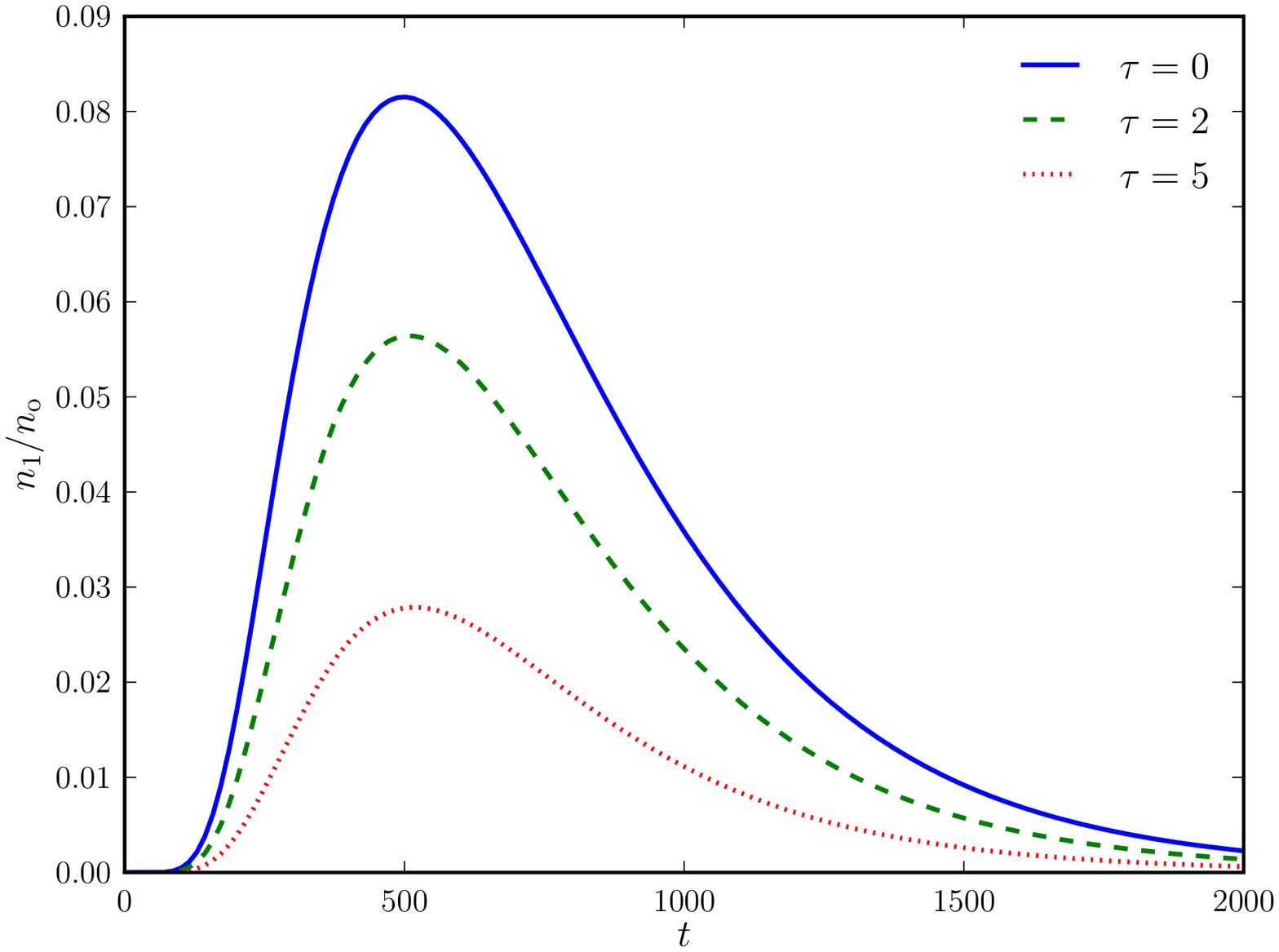}\\
\includegraphics[scale=0.40]{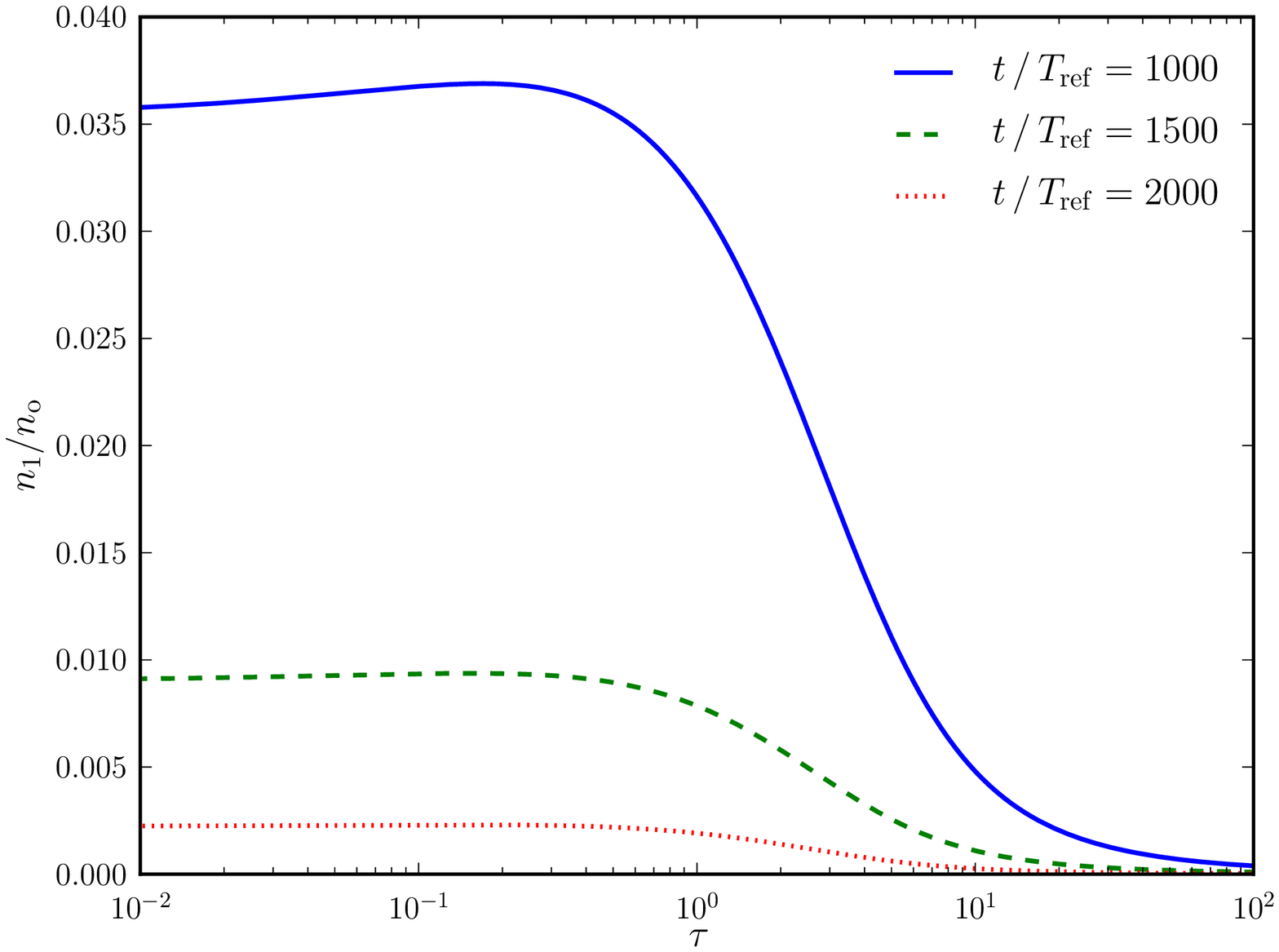}
\end{center}
\caption{Level population of the first excited state, $n_1 / n_{\rm o}$ 
as a function of time in units of $T_{\rm ref}$,
for several reference depths (top panel); and as a function of 
reference depth for several times (bottom panel).  All calculations have been 
performed for the canonical Model III of \S\ref{subsect:Canonical}.}
\label{fig:n1_MIII}
\end{figure}

Figure~\ref{fig:n1_MIII} shows the 
level population of the excited state, $n_1 / n_{\rm o}$ as a function 
of time in units of $T_{\rm ref}$.  Since the upper 
level, $j = 2$, remains negligibly populated throughout the calculation, the 
ground state population is 
$n_0 / n_{\rm o} \approx 1 - n_1 / n_{\rm o}$.  The top panel
shows $n_1 / n_{\rm o}$ as a function of 
$t / T_{\rm ref}$ at several reference depths, $\tau = 0, 2, 5$.
The bottom panel shows $n_1 / n_{\rm o}$ as a function of 
$\tau$ for several times, $t = T_{\rm lc}, 1.5 T_{\rm lc}, 2 T_{\rm lc}$.
The population of the excited state peaks at 
$t / T_{\rm ref} \approx 500$, which is when the radiation field 
close to the $\tau = 0$ boundary has its maximum.  The $j = 0\rightarrow 2$ 
transition, followed by a spontaneous decay $j = 2\rightarrow 1$, is the 
primary mechanism for populating the $j = 1$ level.  Since the radiation 
field in the $j = 0\rightarrow 2$ transition rapidly decreases at line 
centre for $\tau\gg 1$, most of the excitation events occur at 
$\tau \lesssim 1$.  This explains why the population of the excited state 
peaks with the external radiation in the top panel: the light travel time 
to the layers in which significant excitation occurs is relatively small 
compared to the timescales over which variation occurs.  It is interesting 
to note that, as seen in the $t = T_{\rm LC}$ curve in the bottom panel 
of Figure~\ref{fig:n1_MIII}, the population $n_1$ actually has its 
maximum value at $\tau \approx 0.2$.  This is due to the fact that the external 
radiation has not yet been significantly extinguished, while the diffuse 
field is generated by absorption and re-emission into non-zero angles.  
The combination of 
these effects causes the radiation field in the $j = 0\rightarrow 2$ 
transition to peak at $\tau > 0$, leading to a maximum in the $n_1$ population 
at the same point.

\begin{figure}
\begin{center}
\includegraphics[scale=0.40]{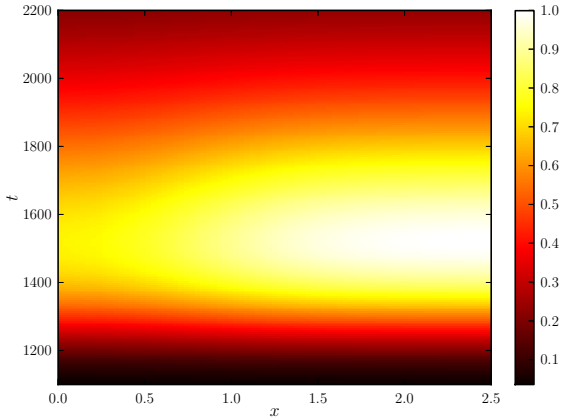}
\end{center}
\caption{Time development of the $u_{\nu}$ line 
profile as a function of $x$.  Results are shown for $\tau = \tau_{\rm max}$ 
and 
$\mu = 1$.  The time is given in units of $t / T_{\rm ref}$, while the 
intensity scale 
is normalized to the maximum value of the radiation field, which occurs  
at $t / T_{\rm ref} \approx 1500$ for $x = 4$.  The calculations were 
performed for 
the canonical Model III of \S\ref{subsect:Canonical}.}
\label{fig:u12full_MIII}
\end{figure}

Figure~\ref{fig:u12full_MIII} shows the time development of the 
$j = 1\rightarrow 2$, $u_{\nu}$ line 
profile as a function of $x$, at $\tau = \tau_{\rm max}$ with $\mu = 1$.
The intensity scale is normalized to the maximum value of the radiation field, 
which occurs at $t / T_{\rm ref} \approx 1500$ for $x = 4$.  At line 
center ($x = 0$), 
the field decreases to its minimum intensity approximately when photons from 
the peak 
emission of the external source cross the medium; as expected, this is when 
the field in 
the line wing ($x = 4$) achieves its maximum intensity.  When the medium 
contains a 
significant fraction of atoms in the $j = 1$ state, absorption at line 
center decreases 
the intensity of the radiation field at earlier times than for frequencies 
$x > 0$.  At $t / T_{\rm ref} > 1800$, when the atoms have depopulated to 
the ground state, 
the time variation of photons at all frequencies is once again the same.

\begin{figure}
\begin{center}
\includegraphics[scale=0.40]{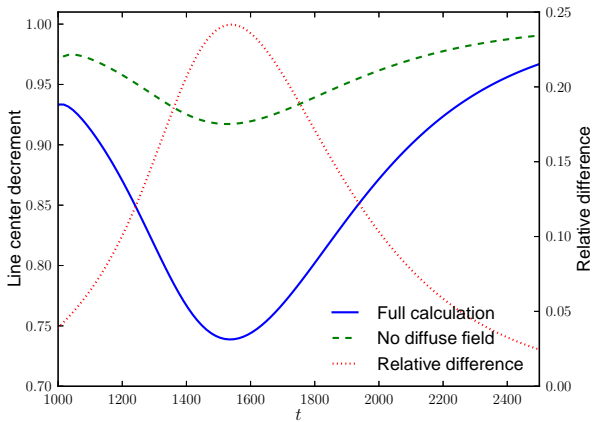}
\end{center}
\caption{Line decrement at $\mu = 1$, $\tau = \tau_{\rm max}$, defined 
as the radiation variable $u_{\nu}$ at line centre divided by its value in 
the line wing ($x = 4$), $u_{\rm cent} / u_{\rm wing}$.  The decrement is shown 
as a function of time in units of $T_{\rm ref}$.  
The solid and dashed curves show results from the full 
calculation and collimated approximation, respectively.  The dotted curve shows 
the relative difference between the full calculation and approximation.  
All calculations have been performed 
for the canonical Model III of \S\ref{subsect:Canonical}.}
\label{fig:u12dec_MIII}
\end{figure}

Figure~\ref{fig:u12dec_MIII} shows $u_{\rm cent} / u_{\rm wing}$ 
for the $j = 1\rightarrow 2$ transition,
at $\tau = \tau_{\rm max}$ and $\mu = 1$, as a function of time (solid curve).   
The decrement can be used to define an effective optical depth, 
\begin{equation}
\label{eq:EffOD}
\tau_{\rm eff} \equiv -\log(u_{\rm cent} / u_{\rm wing}); 
\end{equation}
for the 
canonical Model III, $\tau_{\rm eff}\approx 0.3$.
As expected, the peak effective optical 
depth in the $j = 1\rightarrow 2$ transition occurs at time 
$t / T_{\rm ref} \approx \tau_{\rm max} + 500$, which corresponds to the 
time at which photons from the peak of the external source cross the 
medium.  These photons experience large extinction at line centre due to 
excitation of the atoms into the $j = 1$ level.

Several previous works in the literature have used simplified 
treatments of radiative transfer to infer the state of atomic 
populations in absorbing material around GRBs from the effective optical 
depth in absorption lines of afterglow specra (see \S\ref{sect:Discussion}).  
These works neglected the diffuse field in their solution to 
the radiative transfer equation; 
we refer to this solution as the `collimated approximation' 
\citep[see, e.g.,][]{Vreeswijketal07a, D'Eliaetal09a, Robinsonetal10a}.  
In this case, the external radiation is propagated through 
the slab, but angularly redistributed photons are neglected in determining 
the values of the level populations.
In this approximation, the calculation becomes much simpler, and 
the convergence of the level populations is almost immediate.  The result 
is shown by the dashed curve in Figure~\ref{fig:u12dec_MIII}.  Clearly, 
significant errors are introduced into the calculation of the decrement when 
the 
diffuse component of the radiation field is neglected.  This is due to the 
fact that neglecting the diffuse field implies a lower excitation rate of 
atoms from the ground state to $j = 1$.  If $T_{\rm VR}$ is of order 
$T_{\rm ref}$, then the line centre extinction for photons propagating 
through the medium will be increased when the diffuse field is taken into 
account, resulting in a larger decrement.

\subsection{Variation of Model Parameters 
            (Models IV--VI)}
\label{subsect:Variation}

In this section, we explore the effects of changing the atomic physics; 
the transient amplitude, $A$; and the value of the background 
radiation, $I_{\rm B}$.

\begin{figure}
\begin{center}
\includegraphics[scale=0.40]{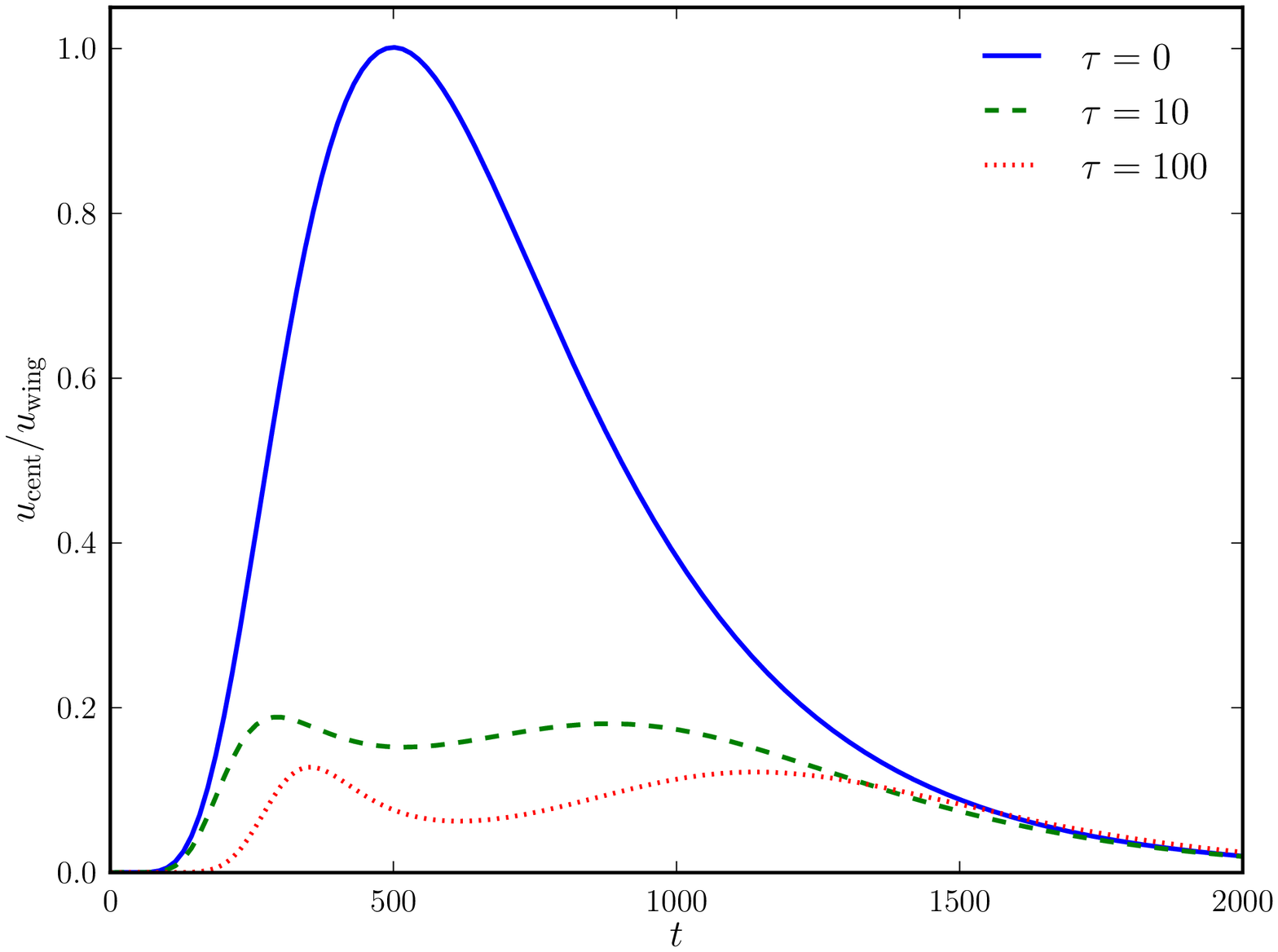}\\
\includegraphics[scale=0.40]{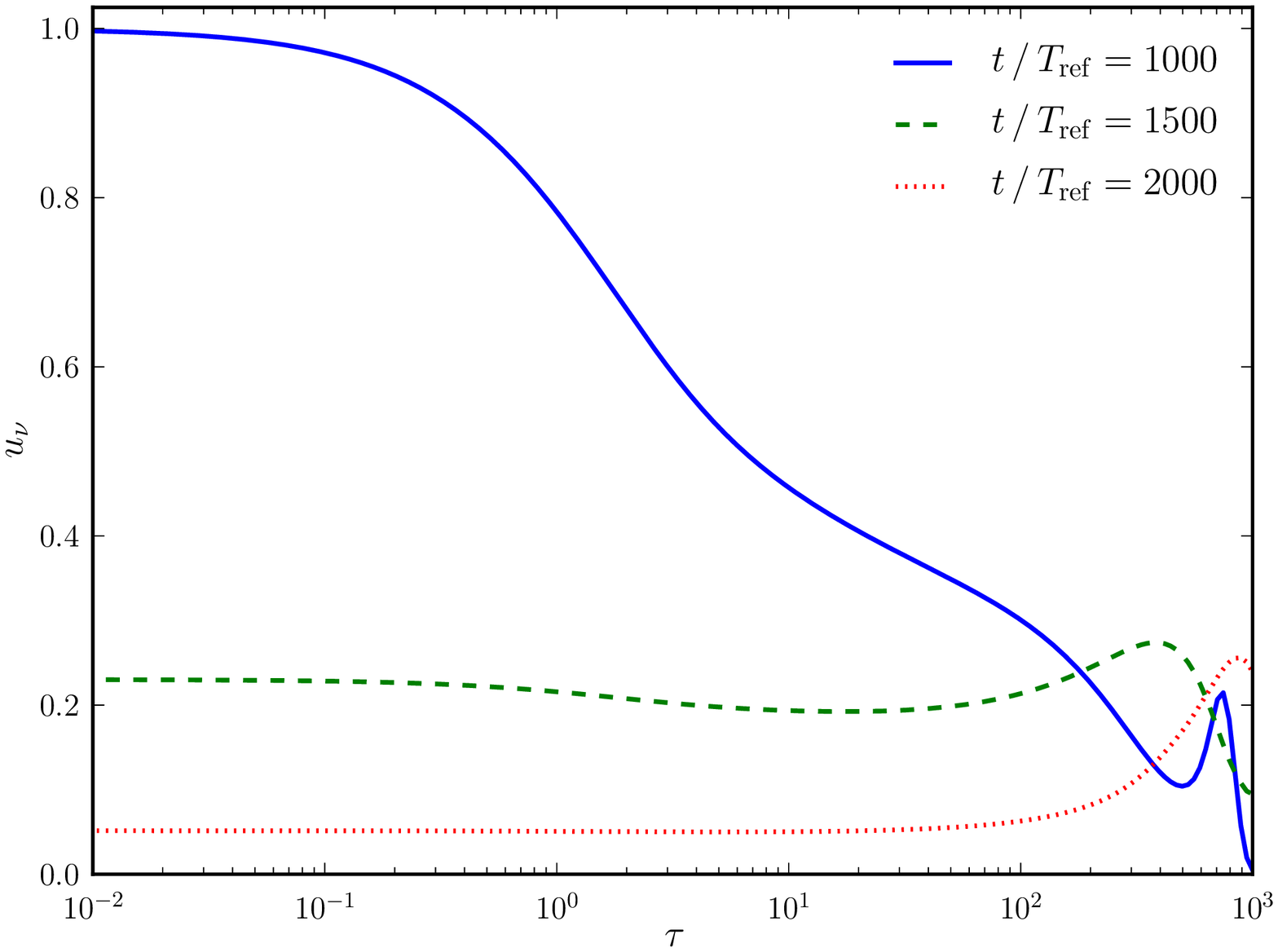}
\end{center}
\caption{Same as Figure~\ref{fig:u12_MIII}, except that results are 
shown for Model IV of \S\ref{subsect:Variation}.} 
\label{fig:u12_MIV}
\end{figure}

We performed a similar calculation 
to that described in Model III, except that we multiplied the 
Einstein coefficients for the $j = 1\rightarrow 2$ transition by a 
factor of ten (Model IV).
Figure~\ref{fig:u12_MIV} is identical to Figure~\ref{fig:u12_MIII} except 
that it shows results for Model IV.
In addition, the results in the top panel were computed for 
$\tau = 0, 10, 100$. 
The figure indicates that 
the effective optical depth for the $j = 1\rightarrow 2$ transition is 
much larger for Model IV than for Model III.  An interesting feature in 
the top panel of Figure~\ref{fig:u12_MIV} is the slight decrease in 
the intensity 
around $t / T_{\rm ref} = \tau + 500$, where the radiation field peaks 
in Figure~\ref{fig:u12_MIII}.  The new feature occurs because of the 
increase in the population of the $j = 1$ level, resulting in larger 
extinction.  Most photons with frequencies near line centre 
of the $j=1\rightarrow 2$ transition that pass through spatial point 
$\tau\approx 500$ at time $t / T_{\rm ref} = \tau + 500$ 
originated in the peak of the external radiation. However, the intensity exhibits 
an overall decrease due to the enhanced extinction.  Such a feature 
was absent from Figure~\ref{fig:u12_MIII} because the total extinction 
was too small to compensate for the increase in intensity from the 
external boundary.  The same phenomenon is responsible for the dip in 
intensity around $\tau\approx 500$ in the solid curve ($t = T_{\rm LC}$) of 
the bottom panel.

\begin{figure}
\begin{center}
\includegraphics[scale=0.40]{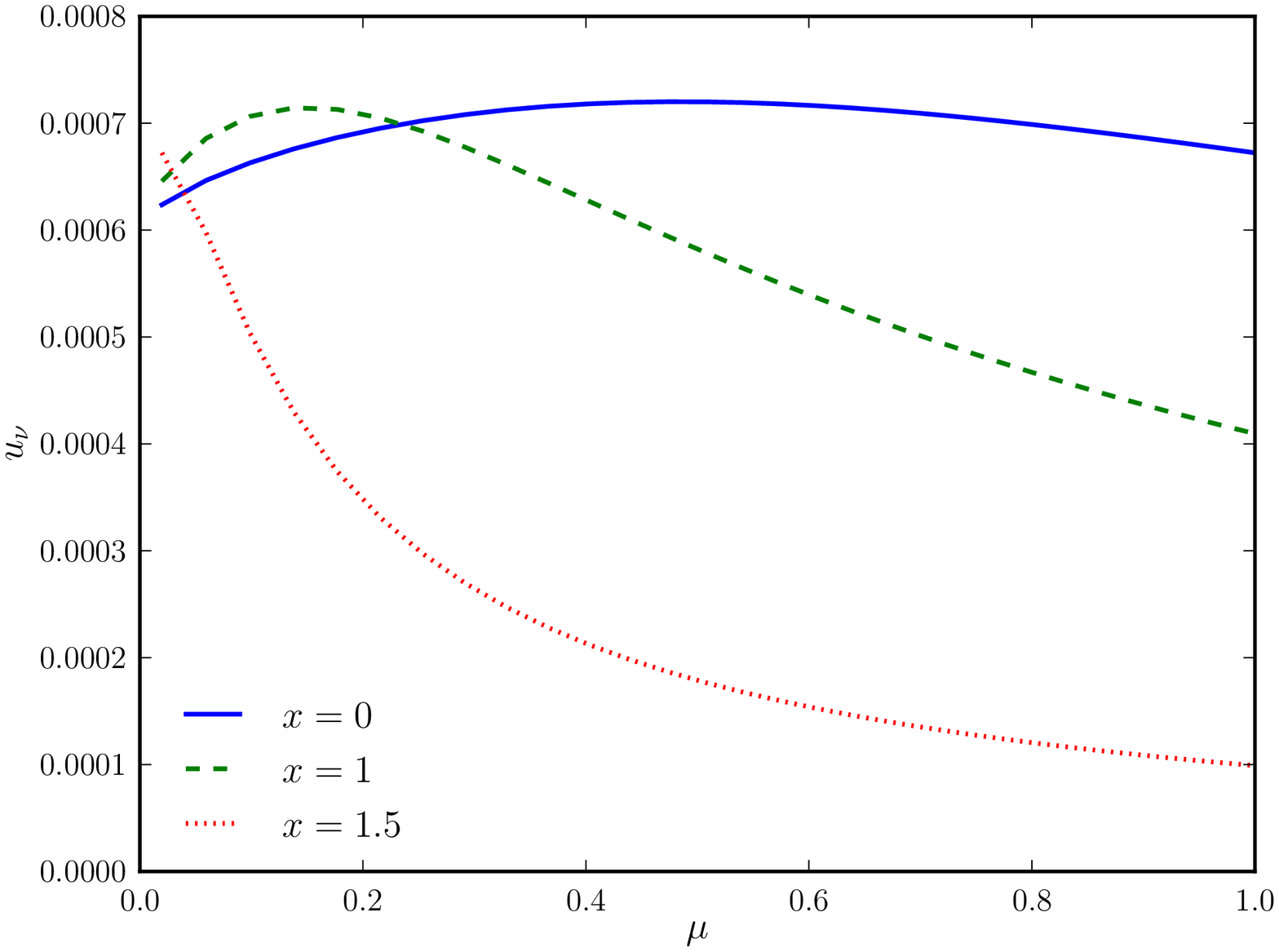}\\
\includegraphics[scale=0.40]{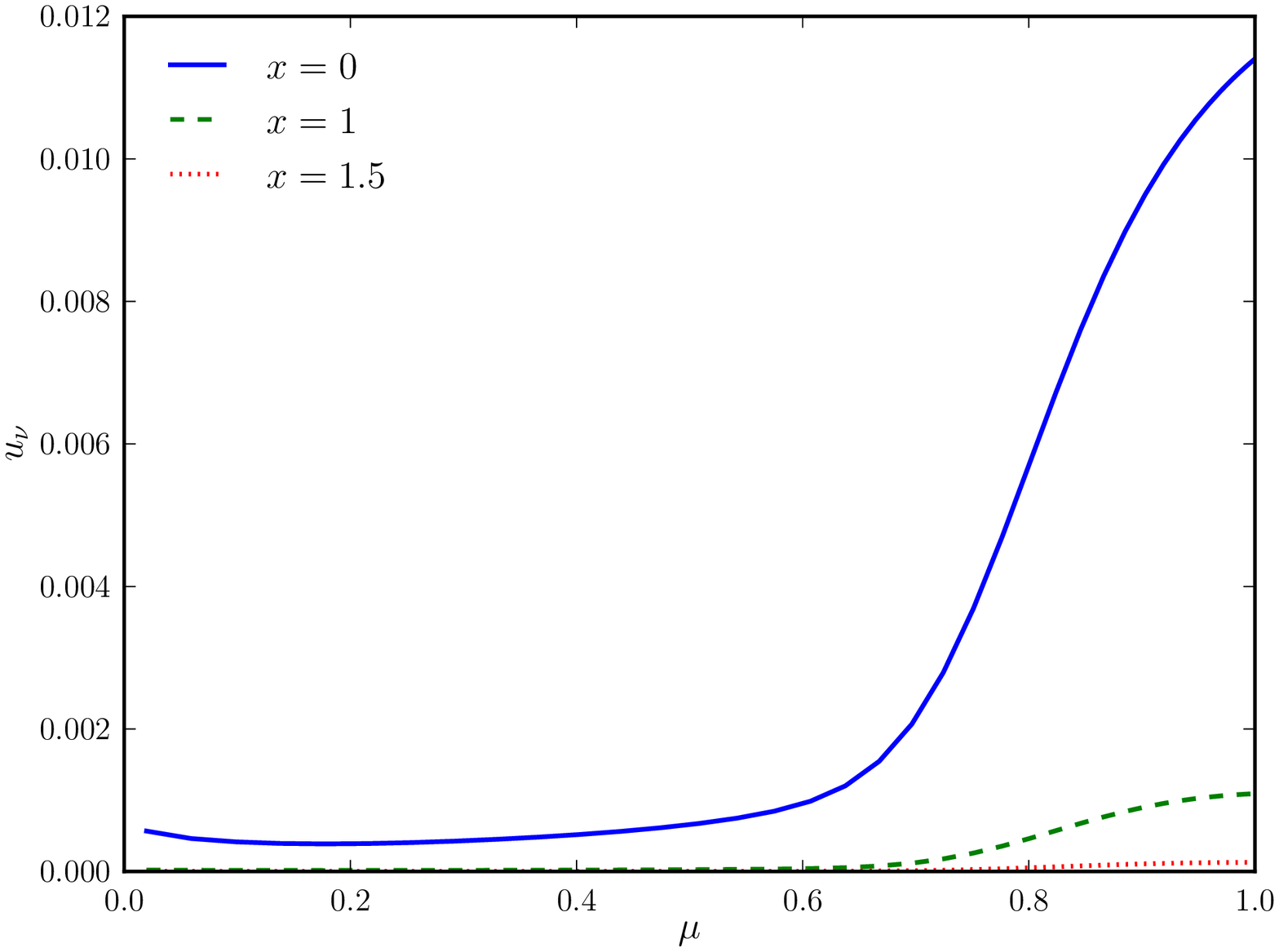}
\end{center}
\caption{Same as Figure~\ref{fig:u12th_MIII}, except that results are shown 
for Model IV of \S\ref{subsect:Variation}.}
\label{fig:u12th_MIV}
\end{figure}

Figure~\ref{fig:u12th_MIV} shows the angular distribution of 
radiation for Model IV.   
The results are similar to those in Figure~\ref{fig:u12th_MIII} for Model III, 
with the radiation field exhibiting increased isotropy at  
$\tau = \tau_{\rm max}$, 
due to increased angular redistribution reflected in the larger value of the 
effective optical depth.

\begin{figure}
\begin{center}
\includegraphics[scale=0.40]{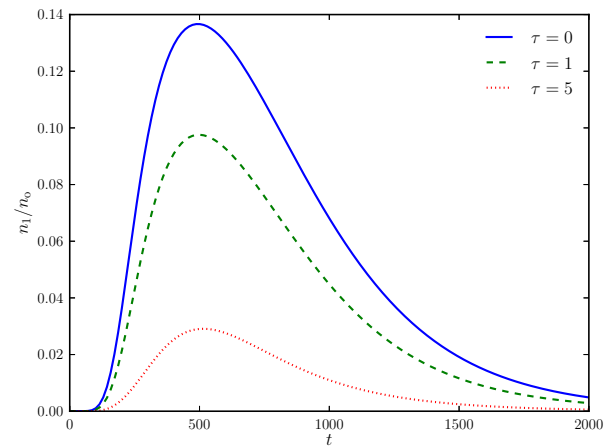}
\end{center}
\caption{Same as the top panel of Figure~\ref{fig:n1_MIV}, except that 
results are shown 
for Model IV of \S\ref{subsect:Variation}.}
\label{fig:n1_MIV}
\end{figure}

Figure~\ref{fig:n1_MIV} is identical to the top panel of 
Figure~\ref{fig:n1_MIII}, except that it shows $n_1 / n_{\rm o}$ as 
a function of time at $\tau = 0, 1, 5$.  At small reference depths, we 
see an increase in the maximum population of level $j = 1$; this is 
due to the increase in the spontaneous emission coefficient for the 
$j = 2\rightarrow 1$ transition.

\begin{figure}
\begin{center}
\includegraphics[scale=0.40]{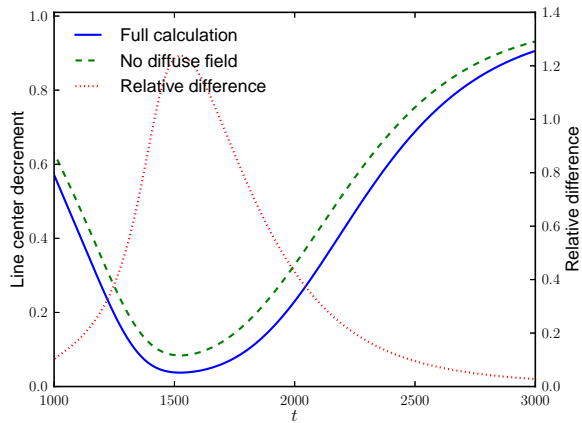}
\end{center}
\caption{Same as Figure~\ref{fig:u12dec_MIII}, except that results are 
shown for Model IV of \S\ref{subsect:Variation}.}
\label{fig:u12dec_MIV}
\end{figure}

Figure~\ref{fig:u12dec_MIV} shows the decrement in the 
$j = 1\rightarrow 2$ transition as a 
function of time in units of $T_{\rm ref}$.  
The plot exhibits the same overall features as 
Figure~\ref{fig:u12dec_MIII}, except that the maximum 
effective optical depth at 
$t / T_{\rm ref} \approx \tau_{\rm max} + 500$ is 
$\tau_{\rm eff} \approx 3$.  
The larger Einstein coefficients 
for the $j = 1\rightarrow 2$ transition cause increased extinction 
per excitation event.
Thus, while both the full calculation and the collimated approximation
exhibit enhanced extinction, there is a larger relative error associated 
with employing the latter (up to 100\%) compared to Model~III.  This is 
due to the fact that there are fewer excitations to the $j = 1$ level in 
the collimated approximation due to the neglect of the diffuse field, 
and this error is amplified in the extinction by the change in atomic 
physics for Model~IV.

We performed another calculation using the same parameters and Einstein 
coefficients as in 
Model III, except for an increase in the amplitude of the transient component 
of equation~\eqref{eq:lightcurve} to $A = 1$ (Model V).

\begin{figure}
\begin{center}
\includegraphics[scale=0.40]{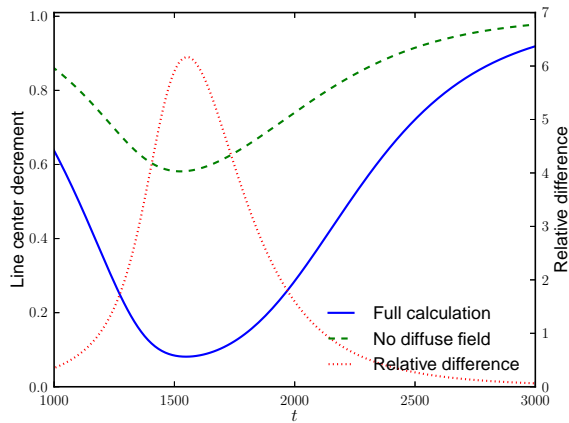}
\end{center}
\caption{Same as Figure~\ref{fig:u12dec_MIII}, except that results are 
shown for Model V of \S\ref{subsect:Variation}.}
\label{fig:u12dec_MV}
\end{figure}

Figure~\ref{fig:u12dec_MV} is identical to Figure~\ref{fig:u12dec_MIV}, 
except that it shows results for Model V.  The maximum effective 
optical depth in the $j = 1\rightarrow 2$ transition is $\tau_{\rm eff} 
\approx 3$.  However, the relative error introduced into the calculation by 
neglecting the diffuse field has increased by a factor of six.  
This is due to the increase in the maximum amplitude of the external 
radiation.  In Model~IV, the increased extinction relative 
to Model~III resulted from the stronger $j = 1\rightarrow 2$ 
transition in the atomic model, which affected both the full calculation 
and the collimated approximation.
In Model~V, the increased 
extinction relative to Model~III is due to enhanced excitation to 
the $j = 1$ level; this affects the full calculation and the 
collimated approximation differently.  In the former, inclusion of the 
diffuse field fully incorporates the effect of increasing the external 
radiation amplitude.  In the latter, a significant fraction of the 
amplitude increase is neglected, leading to a smaller effective 
optical depth for the collimated approximation relative to Model~IV.

We explore a final case in which the background radiation is a 
significant fraction of the transient amplitude (Model VI).  For this case, 
we change the parameters of the canonical Model III to 
$I_{\rm B} / I_{\rm o} = 0.1 $ and $A / I_{\rm o} = 1$.

\begin{figure}
\begin{center}
\includegraphics[scale=0.40]{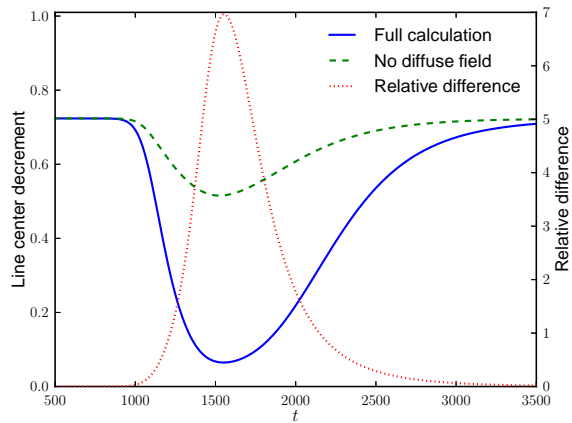}
\end{center}
\caption{Same as Figure~\ref{fig:u12dec_MIII}, except that results are shown
for Model VI of \S\ref{subsect:Variation}.}
\label{fig:u12dec_MVI}
\end{figure}

For the most part, the results for Model VI are similar to those for Model 
V.  This is not unexpected as the maximum amplitude of the 
transient component of the external source is the same in both cases.  One 
major difference is that the system begins and relaxes to an equilibrium state 
in which the diffuse field and excited $j = 1$ level population are non-zero. 
At reference depth $\tau = 0$ for times 
$t / T_{\rm ref} = 0$ and $t / T_{\rm ref} > 10^3$, 
the excited state populations are $n_1 / n_{\rm o} \approx 0.08$ (note that 
this is approximately equal to the maximum population at $\tau = 0$ for 
Model III).  Figure~\ref{fig:u12dec_MVI} shows the line decrement results 
for Model VI.  At $t < T_{\rm LC}$, the line decrement is equal to its 
equilibrium value, $u_{\rm cent} / u_{\rm wing} \approx 0.75$.  After a 
light crossing time, it decreases until the effective optical depth is 
$\tau_{\rm eff} \approx 3$, after which it relaxes back to its 
equilibrium value.

The collimated approximation was altered slightly for the calculations of 
Model~VI.  In this case, the steady-state diffuse field was determined at 
$t / T_{\rm ref} = 0$ and held fixed during the calculation; only the 
collimated radiation was allowed to vary dynamically.
As in Model V, the relative error introduced 
by neglecting changes in the diffuse field is significant, 
greater than $600\%$ at maximum.

\section{Discussion}
\label{sect:Discussion}

We have developed a general formalism for the numerical solution of the TDRTE 
and atomic rate equations using ALI.  We applied our method to a 
three-level atomic 
model that included allowed transitions between the ground and first excited 
states to a common upper level, as well as a forbidden transition between the 
two former states.  Our calculations exhibit convergence of the iterative 
ALI method and, when augmented with additional acceleration 
techniques, can be used in 
efficient calculations of line transfer with realistic atomic models.  Our 
results show time coupled effects in the evolution of the atomic level 
populations, radiation field, and emergent spectra.  We have compared our 
solution to an approximate treatment in which photons that are redistributed 
in angle are neglected (i.e., the diffuse field is ignored), and showed that 
this approximation results in errors of $20 - 600\%$
in the effective optical depth for the $j = 1\rightarrow 2$ transition.
This error increases with the magnitude of $\tau_{\rm eff}$.  
The effect occurs because the diffuse field photons affect 
the level populations and extinction coefficient for the excited 
state transition.  If the time variability of the source falls in the 
intermediate regime (see \S\ref{sect:Introduction}), then photons passing 
through the medium along the line of sight experience increased extinction 
when the diffuse radiation field is included in the calculation.  

These conclusions are relevant for timescales obeying the 
relations $T_{\rm VR} \gtrsim T_{\rm ref}$ and 
$T_{\rm VR} \lesssim T_{\rm LC}$.  Assuming a Doppler line profile in complete 
redistribution, we can write the ratio $T_{\rm VR} / T_{\rm ref}$ as
\begin{equation}
\label{eq:PhysUnits}
\begin{split}
\frac{T_{\rm VR}}{T_{\rm ref}} & = 
	0.893 \left(\frac{T_{\rm VR}}{10^3\text{ s}}\right)
	\left(\frac{10^{-5}}{b/c}\right)\\
	& \times \left(\frac{B_{02}}
		{10^9\text{ cm}^2\text{ s}^{-1}\text{ erg}^{-1}}\right)
	\left(\frac{n_{\rm o}}{\text{cm}^{-3}}\right),
\end{split}
\end{equation}
and the maximum optical depth as
\begin{equation}
\label{eq:taumax}
\begin{split}
\tau_{\rm max} & = 
	2.98 \left(\frac{W}{10^{14}\text{cm}}\right)
	\left(\frac{10^{-5}}{b/c}\right)\\
	& \times \left(\frac{B_{02}}
		{10^9\text{ cm}^2\text{ s}^{-1}\text{ erg}^{-1}}\right)
	\left(\frac{n_{\rm o}}{\text{cm}^{-3}}\right).
\end{split}
\end{equation}
If the timescales characterizing the external radiation source and 
medium satisfy the intermediate regime inequalities, then radiation transfer 
in the astrophysical system should be treated using the ALI methods 
developed in this paper.
If, however, $T_{\rm VR}\ll T_{\rm ref}$,
then the collimated approximation can be used without loss of accuracy.  
This is due to the fact that the diffuse radiation field is
generated on timescales $\gtrsim T_{\rm ref}$, and if the transient
pulse propagates through the medium on smaller timescales, then
the level populations will not be altered by the diffuse field in
time to affect extinction of the external photons.
Conversely, if $T_{\rm VR}\gg T_{\rm LC}$, then the radiation field 
is determined by the steady-state RTE.

Using the scaling of the quantities in equations~\eqref{eq:PhysUnits} 
and \eqref{eq:taumax}, we
can draw conclusions about what kind of astrophysical systems 
require the full TDRTE solution presented in this paper.
In all cases, we assume that the characteristic transition has 
an Einstein coefficient $B_{02}\sim 10^9$~cm$^2$ s$^{-1}$ erg$^{-1}$.

In absorbing systems composed of metal ions with $n_{\rm o} 
\sim 10^{-5}$~cm$^{-3}$, if the width of the medium is $W\gtrsim 1$~pc 
(corresponding to a minimum $\tau_{\rm max}\sim 1$), then the time variability
of the external source must be $T_{\rm VR}\gtrsim 10^8$~s for the 
full ALI solution to apply.
In contrast, 
for absorbing systems at a higher density, $n_{\rm o}\sim 1$~cm$^{-3}$, 
the full ALI solution applies when 
$T_{\rm VR}\sim 10^3$~s and $W\gtrsim 3\times 10^{13}$~cm.

Time variability is common in astrophysical sources.  Notable
examples of highly transient emitters are GRBs, which emit prompt
$\gamma$-ray emission at early times, followed by longer wavelength 
afterglow radiation 
in the X-ray, UV and optical bands.  The afterglow emission occurs 
on timescales that range from minutes to
days.  A large number of studies have examined the time
dependence of afterglow absorption spectra due to photoionization
of the medium; this variability occurs on timescales comparable to the 
observation time
\citep[][]{PernaLoeb98a, Bottcheretal99a, PernaLazzati02a, LazzatiPerna02a, 
Mirabaletal02a, Robinsonetal10a}. 
Observations of time-dependent absorption have been
reported for a number of GRBs
\citep[][]{Thoneetal11a, Vreeswijketal12a, DeCiaetal12a}. 
In a few cases, 
excitation of fine line
transitions of Fe~II were observed in the absence of appreciable
photoionization \citep[][]{Vreeswijketal07a, D'Eliaetal09a}.  
So far, none of the time-dependent radiative transfer calculations 
developed specifically for GRBs has included source terms for 
the diffuse field.
Previous work emphasized the
modelling of time-dependent absorption effects due to photoionization,
rather than the details of line transfer.

Consider the results of our analysis for a GRB 
variability timescale of $T_{\rm LC}\sim 10^3$~s, and typical interstellar
medium hydrogen density of $n_{\rm H}\sim 1$~cm$^{-3}$. 
For metal ions with $n_{\rm o}\sim 10^{-5}$~cm$^{-3}$, the 
condition $T_{\rm VR}\ll T_{\rm ref}$ is satisfied, and the collimated 
approximation is appropriate.
However, for hydrogen, $T_{\rm VR}\sim T_{\rm ref}$, and 
the diffuse field can affect line transfer 
through the absorber, coupling to the photoionization state.  Thus, 
more detailed studies, using the methods described in our paper, should 
be performed when interpreting observations of afterglow line 
variability.  In addition, if the GRB occurs in a molecular
cloud, with densities as high as $\sim 10^5$~cm$^{-3}$, then
a full solution to the time-dependent line transfer problem may be needed
even for metal lines.
  
AGNs form another important class of time variable sources.
Typical AGN soft X-ray variability timescales range from $\sim 1$~ksec
in nearby, small black hole mass Seyfert galaxies, to
$\sim 100$~ksec in distant, massive quasars \citep[][]{Fioreetal98a}.  
The central source is
surrounded by a cloud of absorbers with hydrogen densities as high as
$\sim 10^7-10^8$~cm, located at distances of $\sim
10^{13}-10^{14}$~cm \citep[][]{Baldwinetal95a, Krongoldetal07a}.
If the source variability is about $1$~ksec for an absorber with 
density $n_{\rm o}\sim 10^3$~cm$^{-3}$ (in metal lines), then 
$T_{\rm VR}\sim 10^3\,T_{\rm ref}$.  In this case, the full time-dependent
formulation of line transfer is needed when 
$\tau_{\rm max}\gtrsim 10^3$, implying an absorber width of 
$W\gtrsim 3\times 10^{13}$~cm. 
If the source variability timescale is longer, 
with $T_{\rm VR} \sim 100$~ksec, 
then the full solution is required 
when $\tau_{\rm max}\gtrsim
10^5$, implying an absorber width of $W\gtrsim 3\times 10^{15}$~cm.

The examples above illustrate that, for any 
time variable source, there is a range of absorber properties
for which line transfer must be treated by the ALI method.

In Appendix~\ref{sect:RKMethods}, we further generalize our formalism 
to include a broad class of Runge-Kutta methods for the time 
integrations employed in \S\ref{sect:Numerical}.  These  
techniques will enable future calculations that monitor the 
error in the time grid discretization, or implement an adaptive time step 
algorithm.  Such improvements can 
improve the stability of the radiation field integration, reduce artificial 
oscillations in the solution, and allow more efficient calculations for 
realistic atomic models.

\section*{Acknowledgements}

This work was partially supported by the grant NSF AST-1009396 (RP). 
The authors would like to thank the referee Ivan Hubeny and 
Shane Davis for critical readings of the manuscript and suggestions 
that greatly improved the presentation of our paper.

\bibliographystyle{mn2e}
\bibliography{ms}

\appendix
\section{Runge-Kutta Methods}
\label{sect:RKMethods}

In this Appendix, we outline how 
the iterative scheme of \S\ref{subsect:ALI} can be 
implemented with implicit Runge-Kutta methods 
\citep[see, e.g., \S II.7 of][]{Haireretal91a}.  
For simplicity, we 
neglect collisions in the rate equations and background contributions to 
the emissivity.  Including these terms requires a straightforward 
generalization of the work below.

Applying the method of lines to the simultaneous system of equations 
yields (c.f., \S\ref{subsect:RTESol}):
\begin{align}
\label{eq:NRHS}
f_{j,d} & = \sum_l \bigl(R_{lj,d} n_{l,d} - R_{jl,d} n_{j,d}\bigr),\\
\label{eq:URHS}
\begin{split}
f_{u,di} & = c\chi_{di}\Bigl[\left(v_{d+1/2,i} - v_{d-1/2,i}\right) / 
	\Delta\tau_{di}\\ &\qquad\qquad - u_{di} + S_{di}\Bigr],
\end{split}\\
\label{VRHS}
\begin{split}
f_{v,d+1/2,i} & = c\chi_{d+1/2,i}\Bigl[\left(u_{d+1,i} - u_{di}\right) / 
	\Delta\tau_{di}^+\\&\qquad\qquad\qquad - v_{d+1/2,i}\Bigr].
\end{split}
\end{align}
Note that we used a second order spatial discretization above, 
though more complicated methods are possible.  Higher order methods will 
lead to a more complicated spatial structure for the linear systems 
of equations (see below).  

The Runge-Kutta solution for equation~\eqref{eq:NRHS} can be written
\begin{multline}
\label{NRK}
n_{j,kd} = n_{j,k-1,d}\\ + \Delta t_k \sum_{p=0}^{P-1} b_p 
             f_{j,d}\left(t_{k-1} + 
             c_p\Delta t_k, n_{j,kd}^p\right),
\end{multline}
with
\begin{multline}
\label{NKP}
n_{j,kd}^p = n_{j,k-1,d} + \Delta t_k \sum_{p'=0}^{P-1} a_{pp'} 
             f_{j,d}\Bigl(t_{k-1}\\ + c_p'\Delta t_k, n_{j,kd}^{p'}, 
                          n_{l,kd}^{p'}, \ldots\Bigr),
\end{multline}
where $P$ is the number of stages, and $c_p$, $a_{pp'}, b_p$ are the 
coefficients for the Runge-Kutta method.  The notation 
$n_{j,kd}^{p'}, n_{l,kd}^{p'}, \dots$ indicates that the $p'^{\rm th}$ 
stage quantities for each level are substituted in the expression 
for $f_{j,d}$.
In equations~\eqref{NRK} and \eqref{NKP}, $f_{j,d}$ contains a dependence 
on $u_d$ through the quantities $R_{ld,d}$ defined in \S\ref{subsect:RatesSol}.

We can use the Runge-Kutta solution to equation~\eqref{VRHS} to simplify the 
dependence on $v$ in \eqref{eq:URHS}:
\begin{multline}
\label{VELIM}
v_{k,d+1/2,i}^p = v_{k-1,d+1/2,i} + \Delta t_k \sum_{p'=0}^{P-1} a_{pp'}
   c \chi_{k,d+1/2,i}^{p'}\\ \times \left[\left(u_{k,d+1,i}^{p'} - 
   u_{kdi}^{p'}\right) / 
   \Delta\tau_{kdi}^{p'+} - v_{k,d+1/2,i}^{p'}\right],
\end{multline}
where the quantities $\Delta\tau_{kdi}^{p'+}$ and $\chi_{k,d+1/2,i}^{p'}$ 
are
evaluated at $t_{k-1} + c_{p'}\Delta t_k$ using the values $n_{j,kd}^{p'}$.  
This is a linear system of dimension $P$ in the quantities $v_{k,d+1/2,i}^p$.  
If we define the column vector 
\begin{equation}
{\bf v}_{k,d+1/2,i} = 
\left(v_{k,d+1/2,i}^{p=0}, 
\ldots v_{k,d+1/2,i}^{p=P-1}\right)^T,
\end{equation} 
we can write the system as
\begin{multline}
\label{VLINSYS}
\mathbf{L}_{k,d+1/2,i} \cdot {\bf v}_{k,d+1/2,i} =\\ {\bf M}_{k,d+1/2,i} \cdot 
    \left({\bf u}_{k,d+1,i} - {\bf u}_{kdi}\right) + {\bf v}_{k-1,d+1/2,i},
\end{multline}
where the elements of the $P\times P$ matrices ${\bf L}_{k,d+1/2,i}$ and 
${\bf M}_{k,d+1/2,i}$ are determined by equation~\eqref{VELIM}.  The elements 
of the vector ${\bf v}_{k-1,d+1/2,i}$ are all equal to $v_{k-1,d+1/2,i}$.  
The solution to equation~\eqref{VLINSYS} can be obtained in $\mathcal{O}(P^3)$ 
operations for each triple of indices $(k, d+1/2, i)$.  We write this solution 
as:
\begin{multline}
\label{VFORMAL}
{\bf v}_{k,d+1/2,i} = \boldsymbol{\mathcal{M}}_{k,d+1/2,i}\\
    \cdot \left({\bf u}_{k,d+1,i} - {\bf u}_{kdi}\right) + 
    \boldsymbol{\mathcal{N}}_{k,d+1/2,i}.
\end{multline}

We can substitute equation~\eqref{VFORMAL} into the Runga-Kutta formula for 
$u$ to obtain a formal solution for the diffuse radiation field.  The 
Runge-Kutta solution for $u$ can be written:
\begin{multline}
\label{UKP}
u_{kdi}^p = u_{k-1,di} + \Delta t_k\sum_{p'=0}^{P-1} a_{pp'} 
    c\chi_{kdi}^{p'}\\ \times
    \Bigl[\left(v_{k,d+1/2,i}^{p'} - 
    v_{k,d-1/2,i}^{p'}\right)/\Delta\tau_{kdi}^{p'}\\ - 
    u_{kdi}^{p'} + S_{kdi}^{p'}\Bigr] \Rightarrow
\end{multline}
\begin{multline}
\label{ULINSYSA}
{\bf L}_{kdi} \cdot {\bf u}_{kdi} = {\bf M}_{kdi} \cdot 
   \left({\bf v}_{k,d+1/2,i} 
   - {\bf v}_{k,d-1/2,i}\right)\\ + {\bf N}_{kdi} \cdot 
    {\bf S}_{kdi} + {\bf u}_{k-1,di},
\end{multline}
where the matrices and vectors 
are defined analogously to those in the equation for the $v$ variable. 
If we substitute equation~\eqref{VFORMAL} into \eqref{ULINSYSA}, we obtain 
a system of equations of the form:
\begin{equation}
\label{ULINSYSB}
\begin{split}
-{\bf A}_{kdi} \cdot {\bf u}_{k,d-1,i} + \left({\bf L}_{kdi} + 
{\bf A}_{kdi} + {\bf C}_{kdi}\right)\cdot {\bf u}_{kdi}\\ -
    {\bf C}_{kdi}\cdot {\bf u}_{k,d+1,i} =
    {\bf R}_{kdi} \cdot {\bf S}_{kdi} + {\bf B}_{kdi},
\end{split}
\end{equation}
where the vector ${\bf B}_{kdi}$ contains the dependences on the radiation 
variables at the previous time grid point, $k-1$.
Equation~\eqref{ULINSYSB} can be supplemented by boundary conditions 
connecting the points $d = 0, 1$ and $d = D-2, D-1$. 
These conditions can be derived as in
\S\ref{subsect:RTESol}: the method of lines is applied to first order, 
and the 
time integration proceeds as described above.
Thus, we obtain a block tridiagonal linear system that can be solved in 
$\mathcal{O}\left(P^3 D\right)$ operations for each $(k, i)$.  We 
write the solution to equation~\eqref{ULINSYSB} as
\begin{equation}
\label{FSOLU}
\tilde{\bf u}_{ki} =
\tilde{\boldsymbol{\Lambda}}_{ki} \cdot
\tilde{\bf S}_{ki}
+ \tilde{\bf B}_{ki},
\end{equation}
where $\tilde{\boldsymbol{\Lambda}}_{ki}$ is a $D\times D$ block matrix with 
$P\times P$ matrices as elements, and $\tilde{\bf B}_{ki}$ is a $D\times 1$ 
block vector with $P\times 1$ vectors as elements.

To obtain a formal solution for the diffuse field, we augment our time grid 
with the points $\{t_{k-1} + c_p\Delta t_k\}$.
The solution then consists of the values 
$u_{kdi}^{p=0}, \ldots, u_{kdi}^{p=P-1}$, for $k = 1, \ldots K-1$, where 
the quantities $u_{kdi}^{p=P-1}$ represent the solution on the original $K - 1$ 
time grid points. 
Using the initial conditions, 
we can solve equation~\eqref{FSOLU} successively for $k = 1, \ldots K - 1$.  
The structure of this formal solution in terms of the source functions 
$S_{kd}^p$ is quite complicated, coupling the various Runge-Kutta stages and 
spatial grid points.  However, if we consider the ALI form of the formal 
solution,
\begin{equation}
\label{ALIFSOLU}
u_{kdi}^p = \Lambda_{kdi}^{p*} \left(S_{kdi}^p - S_{kdi}^{p\dag}\right) + 
	u_{kdi}^{p\dag},
\end{equation}
we achieve a significant simplification.  In equation~\eqref{ALIFSOLU}, 
$\Lambda_{kdi}^{p*}$ denotes the diagonal elements of the $P\times P$ matrices 
that are in turn along the diagonal of the $D\times D$ block matrix 
$\tilde{\boldsymbol{\Lambda}}_k$.
The source functions are evaluated with the level populations 
$n_{j,kd}^p, n_{l,kd}^p, \ldots,$ for the current and previous 
$(\dag)$ iterations.
The diagonal elements 
of $\tilde{\boldsymbol{\Lambda}}_{ki}$ can be obtained simultaneously with the 
formal solution, using the block matrix 
form of the equations in Appendix~B of \citet{RybickiHummer91a}. 
As in \S\ref{subsect:ALI}, we use the alternate form of 
the formal solution,
\begin{multline}
\label{ALIPSIFORM}
u_{kdi}^p = \Psi_{kdi}^{p*} \left(\eta_{kdi}^p - \eta_{kdi}^{p\dag}\right) + 
	u_{kdi}^{p\dag}\\ = \Psi_{kdi}^{p*} \sum_{l,j} U_{lj,i} 
	\left(n_{l,kd}^p - n_{l,kd}^{p\dag}\right) + u_{kdi}^{p\dag},
\end{multline}
where $\Psi_{kdi}^{p*} = \Lambda_{kdi}^{p*} / \chi_{kdi}^{p\dag}$.

We can substitute equation~\eqref{ALIPSIFORM} into 
equation~\eqref{NKP} and precondition the resulting non-linear expressions
as described in \S\ref{subsect:RatesSol} and \S\ref{subsect:ALI}.
The ALI method then proceeds as follows: starting with the initial 
conditions, $u_{k=0,di}$, $n_{j,k=0,d}$, we solve for the formal solution 
at $k = 1, \ldots, K - 1$.  At each $k$, we iterate over the populations 
$n_{j,kd}^p$ for all the atomic levels and Runge-Kutta stages, using the 
formal solution for $u_{kd}^{p\dag}$, approximate matrix $\Lambda_{kd}^{p*}$, 
and ALI linear system.  The method is completely analogous to  
the technique described in \S\ref{subsect:ALI}; indeed the equations in 
the main text 
are special cases of Runge-Kutta integrations with coefficients given by 
the Butcher tableaux
$\begin{tabular}{c | c}
1 & 1\\
\hline
& 1
\end{tabular}$ (Backward Euler) and 
$\begin{tabular}{c | cc}
0 & 0 & 0\\
1 & 1/2 & 1/2\\
\hline
& 1/2 & 1/2
\end{tabular}$ (Crank-Nicholson).

\label{lastpage}

\end{document}